\documentclass[10pt,letterpaper,compsoc,conference]{iiswc23}

\usepackage{amsmath,amssymb,amsfonts}
\usepackage{graphicx}
\usepackage[dvipsnames]{xcolor}
\usepackage[final]{microtype}
\usepackage[italic]{mathastext}
\usepackage{libertine}
\usepackage[T1]{fontenc}
\usepackage{textcomp}
\usepackage[varqu,varl]{zi4}
\usepackage[all]{nowidow}
\usepackage[keeplastbox]{flushend}

\usepackage{mathptmx} 
\usepackage[normalem]{ulem}
\usepackage[hyphens]{url}
\usepackage[sort,nocompress]{cite}
\usepackage[bookmarks=true,breaklinks=true,letterpaper=true,colorlinks,linkcolor=black,citecolor=blue,urlcolor=black]{hyperref}
\usepackage{balance}
\usepackage{algorithmic, algorithm}
\usepackage{booktabs, multirow} 
\usepackage{soul}
\usepackage{changepage,threeparttable} 
\usepackage{xspace}
\usepackage{hyperref}
\usepackage{tabulary}
\usepackage{listings}
\usepackage{pifont}

\newcommand{\vrotate}[1]{\rotatebox[origin=c]{90}{#1}}

\newcommand{\libname}[1]{\textbf{\texttt{#1}}}

\newcommand{\cmark}{\ding{51}}%
\newcommand{\xmark}{\ding{55}}%

\usepackage{xcolor}
\usepackage{listings}

\definecolor{mGreen}{rgb}{0,0.6,0}
\definecolor{mGray}{rgb}{0.5,0.5,0.5}
\definecolor{mPurple}{rgb}{0.58,0,0.82}
\definecolor{backgroundColour}{rgb}{0.95,0.95,0.92}
\definecolor{mygreen}{RGB}{0, 100, 30}
\definecolor{myred}{RGB}{200, 00, 00}

\lstdefinestyle{CStyle}{
    backgroundcolor=\color{backgroundColour},   
    commentstyle=\color{mGreen},
    keywordstyle=\color{magenta},
    numberstyle=\tiny\color{mGray},
    stringstyle=\color{mPurple},
    basicstyle=\footnotesize,
    breakatwhitespace=false,         
    breaklines=true,                 
    captionpos=b,                    
    keepspaces=true,                 
    numbers=left,                    
    numbersep=5pt,                  
    showspaces=false,                
    showstringspaces=false,
    showtabs=false,                  
    tabsize=2,
    language=C
}

\lstdefinestyle{bash}{%
  backgroundcolor=\color{yellow!10},%
  basicstyle=\small\ttfamily,%
  identifierstyle=\color{black},%
  keywordstyle=\color{blue},%
  keywordstyle={[2]\color{cyan}},%
  keywordstyle={[3]\color{olive}},%
  stringstyle=\color{teal},%
  commentstyle=\itshape\color{orange},%
}%

\newcommand{\swan}[1]{Swan}
\newcommand{\ignore}[1]{}

\title{Vector-Processing for Mobile Devices: Benchmark and Analysis}

\makeatletter
\newcommand{\linebreakand}{%
  \end{@IEEEauthorhalign}
  \hfill\mbox{}\par
  \mbox{}\hfill\begin{@IEEEauthorhalign}
}
\makeatother

\makeatletter
\newcommand*\titleheader[1]{\gdef\@titleheader{#1}}
\AtBeginDocument{%
  \let\st@red@title\@title%
  \def\@title{%
    \vskip-3.0em \bgroup\normalfont\normalsize\centering\@titleheader\par\egroup
    \vskip1.0em\st@red@title}
}
\makeatother

\author{\IEEEauthorblockN{Alireza Khadem\IEEEauthorrefmark{2}}
    \IEEEauthorblockA{\href{mailto:arkhadem@umich.edu}{arkhadem@umich.edu}}
    \and
    \IEEEauthorblockN{Daichi Fujiki\IEEEauthorrefmark{3}}
    \IEEEauthorblockA{\href{mailto:dfujiki@keio.jp}{dfujiki@keio.jp}}
    \and
    \IEEEauthorblockN{Nishil Talati\IEEEauthorrefmark{2}}
    \IEEEauthorblockA{\href{mailto:talatin@umich.edu}{talatin@umich.edu}}
    \and
    \IEEEauthorblockN{Scott Mahlke\IEEEauthorrefmark{2}\IEEEauthorrefmark{4}}
    \IEEEauthorblockA{\href{mailto:mahlke@umich.edu}{mahlke@umich.edu}}
    \and
    \IEEEauthorblockN{Reetuparna Das\IEEEauthorrefmark{2}}
    \IEEEauthorblockA{\href{mailto:reetudas@umich.edu}{reetudas@umich.edu}}
    \linebreakand
    \IEEEauthorblockA{\textit{\IEEEauthorrefmark{2}University of Michigan, \IEEEauthorrefmark{3}Keio University, \IEEEauthorrefmark{4}Nvidia Research}}
    }

\begin{document}

\pagenumbering{arabic}







\titleheader{In Proceedings of the IEEE International Symposium on Workload Characterization (IISWC), Oct. 2023 \\ Copyright IEEE, 2023}

\maketitle
\thispagestyle{plain}
\pagestyle{plain}

\begin{abstract}
Vector processing has become commonplace in today's CPU microarchitectures.
Vector instructions improve performance and energy which is crucial for resource-constraint mobile devices.
The research community currently lacks a comprehensive benchmark suite to study the benefits of vector processing for mobile devices.
This paper presents \swan{}---an extensive vector processing benchmark suite for mobile applications.
\swan{} consists of a diverse set of data-parallel workloads from four commonly used mobile applications: operating system, web browser, audio/video messaging application, and PDF rendering engine.
Using \swan{} benchmark suite, we conduct a detailed analysis of the performance, power, and energy consumption of vectorized workloads, and show that:
(a) Vectorized kernels increase the pressure on cache hierarchy due to the higher rate of memory requests.
(b) Vector processing is more beneficial for workloads with lower precision operations and higher cache hit rates.
(c) Limited Instruction-Level Parallelism and strided memory accesses to multi-dimensional data structures prevent vector processing benefits from scaling with more SIMD functional units and wider registers.
(d) Despite lower computation throughput than domain-specific accelerators, such as GPU, vector processing outperforms these accelerators for kernels with lower operation counts.
Finally, we show five common computation patterns in mobile data-parallel workloads that dominate the execution time.
\end{abstract}
\section{Introduction}

\begin{table*}[ht]\centering
\caption{Prior General-Purpose and Domain-Specific Benchmark Suites for Mobile Platforms.}
\label{tab:benchmarks}
\scriptsize
\addtolength{\tabcolsep}{-1pt}
\begin{tabular}{l|cccccc}\toprule
\textbf{Benchmark Suite} &
\textbf{Microarchitecture} &
\textbf{Implementation} &
\textbf{Application Domain} &
\textbf{Metrics} &
\textbf{Diversity Analysis} &
\textbf{Open Source}  \\\midrule
EEMBC CoreMark~\cite{coremark} &
Single-core &
Scalar &
\textcolor{mygreen}{Mobile} &
\textcolor{mygreen}{Perf./Pow.} &
\xmark &
\textcolor{mygreen}{\cmark} \\

Geekbench~\cite{geekbench} &
Single-core &
Scalar &
Desktop/Server/\textcolor{mygreen}{Mobile} &
Perf &
\textcolor{mygreen}{\cmark} &
\xmark \\
EEMBC MultiBench~\cite{multibench} &
CMP &
UNK &
\textcolor{mygreen}{Mobile} &
Perf. &
\xmark & 
\xmark \\
ParVec~\cite{parvec} &
\textcolor{mygreen}{Vector} (CMP)$^{*}$ & Pthread/AVX/\textcolor{mygreen}{Neon} &
Desktop/Server &
\textcolor{mygreen}{Perf./Pow.} &
\textcolor{mygreen}{\cmark} &
\textcolor{mygreen}{\cmark} \\

RiVEC~\cite{rivec} &
\textcolor{mygreen}{Vector} (CMP/GPU)$^{*}$ &
RVV &
Desktop/Server &
Perf. &
\textcolor{mygreen}{\cmark} &
\textcolor{mygreen}{\cmark} \\

VectorBench~\cite{vectorbench} &
\textcolor{mygreen}{Vector} &
AVX/\textcolor{mygreen}{Neon} &
Desktop/Server &
Perf. &
\xmark &
\textcolor{mygreen}{\cmark} \\

VComputeBench~\cite{vcomputebench} &
\textcolor{mygreen}{Vector} (GPU)$^{*}$ &
Vulkan &
\textcolor{mygreen}{Mobile} &
Perf. &
\textcolor{mygreen}{\cmark} &
\textcolor{mygreen}{\cmark} \\

MEVBench~\cite{mevbench}$^{+}$ &
CMP &
Pthread/OpenCV &
\textcolor{mygreen}{Mobile} &
Perf. &
\xmark &
\textcolor{mygreen}{\cmark} \\

MLPerf~\cite{janapa2022mlperf}$^{+}$ &
CMP/GPU/DSP &
VendorSDK/TFLite &
\textcolor{mygreen}{Mobile} &
Perf. &
\xmark &
\textcolor{mygreen}{\cmark} \\

ARBench~\cite{arbench}$^{+}$ &
CMP/GPU/DSP &
ARCore/OpenGL &
\textcolor{mygreen}{Mobile} &
Perf. &
\xmark &
\textcolor{mygreen}{\cmark} \\

WiBench~\cite{wibench}$^{+}$ &
\textcolor{mygreen}{Vector} &
SSE &
Desktop/\textcolor{mygreen}{Mobile} &
Perf. &
\xmark &
\textcolor{mygreen}{\cmark} \\

\textbf{\swan{}} (this work) &
\textcolor{mygreen}{Vector} &
\textcolor{mygreen}{Neon} &
\textcolor{mygreen}{Mobile} &
\textcolor{mygreen}{Perf./Pow.} &
\textcolor{mygreen}{\cmark} &
\textcolor{mygreen}{\cmark} \\
\bottomrule
\multicolumn{2}{l}{$^+$ Domain-Specific Benchmark.}&
\multicolumn{5}{l}{$^*$Benchmark redevelops workloads of other benchmark suites (features of the source benchmark).}

\end{tabular}
\vspace{-5mm}
\end{table*}

Vector processing was first introduced in supercomputers~\cite{tiasc,cscstar,cray1} and has become commonplace in today's CPU microarchitectures.
Both
server-grade and resource-limited mobile CPUs integrate vector pipelines.
Vector Instruction Set Architecture (ISA) extensions efficiently encode multiple operations within a single instruction and enjoy fine-grain scalar-vector instruction interleaving~\cite{vectorarchitecture}.
The performance and energy improvement of vector processing is even more significant to the resource-limited mobile processors.
This work aims to study the strengths and limitations of vector processing for mobile devices and provides a comprehensive benchmark suite for further research in this area.

Despite the availability of several domain-specific and general-purpose mobile benchmark suits~\cite{parvec,rivec,vectorbench,vcomputebench,mevbench,janapa2022mlperf,arbench,wibench}, a comprehensive benchmark suite for evaluating vector processing in mobile devices is lacking.
This is because prior vector benchmark suites include Chip Multiprocessor (CMP) or GPU applications for vector processing evaluation, where the control-flow divergence limit the power of vector processing.
These applications are offloaded to domain-specific accelerators of the mobile SoCs.
In addition, these benchmark suites contain desktop and server applications with massive parallelism that are offloaded to the cloud-computing in mobile applications.
Therefore, providing a benchmark suite for mobile applications significantly eases the state-of-the-art research in this area.

This paper introduces \textbf{\swan{}}---a benchmark suite for studying vector processing in mobile devices.
\swan{} encompasses vectorized workloads for resource-constraint mobile processing platforms from multimedia processing, graphics, data compression, cryptography, and string utility domains.
We assemble a diverse set of optimized mobile applications and conduct an in-depth performance and energy analysis study.
\swan{} benchmark suite is maintained online at \href{https://github.com/arkhadem/Swan}{https://github.com/arkhadem/Swan}.

To create \swan{}, we extensively study the source code of four widely-used mobile applications: Chromium~\cite{chromium} (web browsing), Android~\cite{android} (operating system), WebRTC~\cite{webrtc} (audio/video messaging service), and PDFium~\cite{pdfium} (PDF rendering engine).
We carefully choose 12 libraries that are commonly used among these applications to cover a large set of workloads representative of mobile applications.
We select a rich set of 59 data-parallel kernels, each of which executes an independent task.
We study the vector instruction set and instruction reduction diversity of these workloads.

Using the \swan{} benchmark suite, we conduct a detailed study of vector processing for mobile applications, specifically Arm Neon architecture, in terms of performance and energy/power consumption.
We show that vector processing is more beneficial for workloads with lower operation precision and higher cache hit rate.
In addition, we illustrate the limited performance improvement of compiler auto-vectorization for the scalar implementation since the scalar code is optimized for superscalar execution.
Our analysis shows substantial energy savings of vector processing due to significantly lower execution time.
On the other hand, vector processing increases chip power consumption in workloads with large working set sizes due to the higher main memory access rates.
Studying three different core microarchitectures of a big.LITTLE Arm architecture proves that more vector pipelines are not always beneficial due to the limited vector Instruction Level Parallelism (ILP).


We introduce five common computation patterns of vector implementations: reduction, random memory access, strided memory access, matrix transposition, and vector APIs.
We show the importance of these computation patterns in terms of the number of workloads exhibiting them, and the average fraction of workload instructions they consume.

Arm Neon limits vector register width to 128 bits and our evaluated mobile SoC contains only two vector execution pipelines.
To further study the scalability of vector processing with wider registers and more vector pipelines, we implement a fake Arm Neon library (available on \href{https://github.com/arkhadem/Swan}{GitHub}) and re-develop eight representative workloads with wider vector registers up to 1024 bits using an in-house cycle-accurate simulator.
We show that wider vector registers provide substantial speedup for workloads with streaming memory access patterns.
When Data-Level Parallelism (DLP) is not evenly divisible by the register width or strided memory accesses are needed to access a multidimensional data structure, vector engine utilization decreases and workloads hardly enjoy wider implementations.
Besides, high register pressure limits Instruction Level Parallelism (ILP), which in turn drops the utilization of vector pipelines and prevents performance from scaling with more vector execution pipelines.

Finally, we show the limitations of domain-specific accelerators (GPU and DSP) in terms of data transfer and kernel launch overheads.
We compare GPU and vector processing performance for various problem sizes, showing that GPU, despite having higher throughput, only outperforms Neon in workloads with high operation counts.

\swan{} is the \textbf{first} benchmark suite for mobile vector processing.
Key contributions of this work are:
(a) \swan{}, an open-source and true vector processing benchmark suite with 59 diverse data-parallel kernels of four commonly-used mobile applications.
(b) Performance, power, and energy analysis of mobile vector processing benefits and compiler vectorization limitations.
(c) Performance scalability study of vector processing with wider instructions and more vector execution pipelines.
(d) Evaluating domain-specific acceleration bottlenecks for fine-grain data-parallel workloads.

\ignore{
\begin{itemize}
    \setlength{\parskip}{0mm} 
    \setlength{\itemsep}{1mm} 
    
    \item \swan{}, an open-source and true vector processing benchmark suite with 59 diverse data-parallel kernels of four commonly-used mobile applications.
    
    \item Performance, power, and energy analysis of mobile vector processing benefits and compiler vectorization limitations.
    
    \item Performance scalability study of vector processing with wider instructions using in-house simulation infrastructure.
    
    \item Evaluating domain-specific acceleration bottlenecks for fine-grain data-parallel workloads.
    
\end{itemize}
}
\vspace{-2mm}

\section{Background and Motivation}

\subsection{Vector-Processing}~\label{sec:vector_processing}

\vspace{-2mm}

Vector Processing was initially employed in supercomputers~\cite{tiasc,cscstar,cray1} to accelerate workloads with massive parallelism~\cite{vectorarchitecture}.
Vector Instruction Set Architecture (ISA) extension reduces the number of instructions by encoding multiple operations, improving performance and energy consumption.
Due to the instruction efficiency, general-purpose microprocessors employed Vector ISA extensions for multimedia processing workloads with high DLP~\cite{ultrasparc,parisc,mmxadoption}.
Today, various ISAs provide vector extensions that exploit fine-grain DLP in a broad set of applications.

Vector processing plays a crucial role in mobile devices with limited resources due to performance and energy efficiency. 
Arm first employed vector processing by Vector Floating-point (VFP) Coprocessor~\cite{vfp11}.
State-of-the-art Arm architectures integrate vector execution pipelines into mobile processors with Advanced-SIMD (ASIMD) functional units.
While modern mobile SoCs incorporate various domain-specific accelerators, vector processing enjoys tight integration with the scalar pipeline, which is especially important for kernels with \textit{fine-grain scalar-vector instruction interleaving}.

\subsection{Benchmark Suites}~\label{sec:benchmark_suites}

\vspace{-3mm}

Benchmarks ease the evaluation of instruction sets, compiler optimizations, and architecture bottlenecks.
Table~\ref{tab:benchmarks} shows various general-purpose and domain-specific benchmark suites for mobile processors.
We observe that the research community lacks a benchmark suite containing different state-of-the-art vector processing applications for mobile devices.
In this section, we describe the design goals of \swan{} benchmark suite and discuss the opportunities for improvement in prior works within each aspect.


\textbf{Target Microarchitecture and Implementation.}
Different benchmark suites such as Embedded Microprocessor Benchmark Consortium (EEMBC) MultiBench~\cite{multibench} and Geekbench~\cite{geekbench} focus on the Single-Core scalar performance of mobile processors.
Due to the lack of enough parallelism, it is reported that these benchmark suites face difficulty exercising vector engines~\cite{geekbench}. 
EEMBC MultiBench~\cite{multibench} contains Chip Multiprocessor (CMP) workloads.
In addition, various benchmark suites are available for GPUs~\cite{parboil,rodinia,polybench}.
However, Task-Level Parallelism (TLP) of CMP and GPU workloads is not suitable for vector processing.

\textbf{Application Domain.}
Desktop and Server workloads encompass extensive problem sizes and exhibit greater DLP levels than mobile applications.
For example, RiVEC~\cite{rivec} and ParVec~\cite{parvec} contain Financial Analysis and Data Mining workloads.
Mobile applications offload these workloads to cloud computing due to the huge problem size. In addition, various applications that mobile vector processors executed in the past are now offloaded to domain-specific accelerators.
For example, VectorBench~\cite{vectorbench} and ParVec~\cite{parvec} provide video processing workloads that are efficiently offloaded to mobile GPUs. MEVBench~\cite{mevbench}, MLPerf~\cite{janapa2022mlperf}, ARBench~\cite{arbench}, and WiBench~\cite{wibench} are domain-specific benchmark suites that contain a single workload, such as Computer Vision, Machine Learning, Augmented/Virtual Reality, and Wireless Signal Processing workloads, respectively.

\textbf{Metrics, Diversity Analysis, and Open Source.}
Prior vector benchmark suites~\cite{rivec,vectorbench} lack Power and Energy analysis that is crucial to resource-limited mobile devices.
Instruction diversity is yet another important factor for benchmark design to exercise different aspects of the design.

Despite prior works, \swan{} encompasses \textit{vector} workloads from a wide set of \textit{mobile} applications. Section~\ref{sec:patterns} shows that \swan{} benchmarks contain different computation patterns and Section~\ref{sec:appprocessor} studies the inadequacy of domain-specific accelerators for these benchmarks.
In Section~\ref{sec:perfenergy}, we study mobile platform-specific performance metrics, auto-vectorization challenges, and instruction diversity of the benchmarks.
We release \swan{} as an open-source benchmark suite on \href{https://github.com/arkhadem/Swan}{GitHub} to ease the study of vector ISA, compiler, and architecture.
\vspace{-2mm}

\section{The \swan{} Benchmark Suite}~\label{sec:benchmark}

\vspace{-6mm}


\subsection{Scope of Benchmark}

To create \swan{} benchmark suite, we target real-world \textbf{mobile applications}, analyze their common libraries, and select time-consuming data-parallel kernels.
Table~\ref{tab:libraries} shows these libraries and their usage across four mobile applications:
(a) \textit{Chromium Project}~\cite{chromium} encompass Chromium and Chromium OS, which contain the source code of Google Chrome browser and Google Chrome OS.
(b) \textit{Android Project}~\cite{android} contains the source code and scripts for Android operating system.
(c) \textit{WebRTC Project}~\cite{webrtc} provides Real-Time (voice/text/video) Communication (RTC) APIs for many messaging and audio/video conferencing platforms such as Zoom, Microsoft Teams, Slack, or Google Meet.
And (d) \textit{PDFium}~\cite{pdfium} is a PDF rendering engine in Google Chrome and Microsoft Edge browsers.

Our code analysis shows 12 common \textbf{libraries} between these applications that contain data-parallel kernels and Arm Neon implementations.
We profile Chrome execution with GPU acceleration while browsing top visited websites~\cite{topvisited} and choose nine libraries that are not offloaded to GPU due to the offloading overheads (Section~\ref{sec:appprocessor}).
Table~\ref{tab:libraries} shows the maximum and average execution time of Chrome consumed by these libraries.
\swan{} also encompasses three traditional vector processing application domains, \textit{i.e.}, audio and video codecs and Machine Learning for low-end mobile products that lack GPU or DSP due to area and power constraints.
Note that \swan{} is a benchmark suite for CPU vector processing exploration and does not provide GPU and DSP applications.

We carefully separate and benchmark 59 \textbf{data-parallel kernels} from libraries, each of which performs an independent task or algorithm. We prioritize algorithms that are not specific to a library.
For example, \libname{libvpx} contains many data-parallel kernels for VP8 and VP9 video codecs, but we choose a subset of four kernels that are commonly used in the coding process of different video codecs.

\begingroup
\setlength{\tabcolsep}{3pt} 
\renewcommand{\arraystretch}{1} 
\begin{table}[t]
\scriptsize
  \renewcommand\arraystretch{1}
  \centering
  \caption{Accelerated Libraries: Domain, Usage, and Chromium Exe. Time.}
  \scalebox{1}{
  \begin{tabular}{|c|c|c||c|c|c|c||c|c|}
    \hline
    \multirow{5}{*}{\textbf{Library}} &
    \multirow{5}{*}{\textbf{Domain}} &
    \multirow{5}{*}{\vrotate{\textbf{Symbol}}} &
    \multirow{5}{*}{\vrotate{\textbf{Chromium}}} &
    \multirow{5}{*}{\vrotate{\textbf{Android}}} &
    \multirow{5}{*}{\vrotate{\textbf{WebRTC}}} &
    \multirow{5}{*}{\vrotate{\textbf{PDFium}}} &
    \multirow{5}{*}{\vrotate{\textbf{Max. (\%)}}} &
    \multirow{5}{*}{\vrotate{\textbf{Avg. (\%)}}} \\
    &&&&&&&& \\
    &&&&&&&& \\
    &&&&&&&& \\
    &&&&&&&& \\
    \hline \hline
    \libname{libjpeg-turbo} & Image Processing & \libname{LJ} & \cmark & \xmark & \xmark & \cmark & 6.8 & 2.4 \\
    \hline
    \libname{libpng} & Image Processing & \libname{LP} & \cmark & \xmark & \xmark & \cmark & 0.8 & 0.3 \\
    \hline
    \libname{libwebp} & Image Processing & \libname{LW} & \cmark & \xmark &\xmark  & \cmark & 7.3 & 1.7  \\
    \hline
    \libname{Skia} & Graphics & \libname{SK} & \cmark & \cmark & \xmark & \cmark & 8.5 & 4.6 \\
    \hline
    \libname{WebAudio} & Audio Processing & \libname{WA} & \cmark & \xmark & \cmark & \xmark & 16.3 & 2.5 \\
    \hline
    \libname{PFFFT} & Audio Processing & \libname{PF} & \cmark & \cmark & \cmark & \xmark & 5.6 & 1.3 \\
    \hline
    \libname{zlib} & Data Compression & \libname{ZL} & \cmark & \cmark & \xmark & \cmark & 0.4 & 0.2\\
    \hline
    \libname{boringssl} & Cryptography & \libname{BS} & \cmark & \cmark & \cmark & \xmark & 0.9 & 0.6 \\
    \hline
    \libname{Opt. Routines} & String Utilities & \libname{OR} & \cmark & \cmark & \cmark & \cmark & 9.6 & 1.2 \\
    \hline
    \libname{libopus} & Audio Processing & \libname{LO} & \cmark & \cmark & \cmark & \xmark & - & - \\
    \hline
    \libname{libvpx} & Video Processing & \libname{LV} & \cmark & \cmark & \cmark & \xmark & - & - \\
    \hline
    \libname{XNNPACK} & Machine Learning & \libname{XP} & \cmark & \cmark & \xmark & \xmark & - & - \\
    \hline
   \end{tabular}
   }
  \label{tab:libraries}
 \vspace{-6mm}
\end{table}
\endgroup

\subsection{Workloads}~\label{sec:workloads}
Table~\ref{tab:libraries} shows the 12 libraries of \swan{} benchmark suite.

Image processing is a primary application for vector processing.
\libname{libjpeg-turbo}, \libname{libpng}, and \libname{libwebp} are all image processing libraries used in Chromium and PDFium projects for JPEG, PNG, and WEBP image codecs.
\libname{Skia} is a graphics library that provides rasterization (\textit{i.e.}, converting paint operations to pixel bitmap) APIs as a graphics engine for other mobile applications.
These libraries provide fine-grain APIs that process a block, row, channel, or multiple channels of the image.
These APIs:
(1) modify the color space (Gray, RGBA, and YCbCR), color code (PNG's true and indexed color), or size (down/upsample or vertical/horizontal convolution) of the image,
(2) apply prediction filters (WEBP's DC, TrueMotion, Vertical, and Horizontal predictions) for image (de)compression,
or (3) enhance image quality (\libname{libwebp}'s Sharp YUV filters).
Due to the fine-grain interleaving of scalar and vector APIs, domain-specific acceleration of these image processing libraries imposes a non-negligible cost of kernel launch overhead.
In addition, while \libname{Skia} exploits GPU for accelerated rasterization, different operations that convert data to GPU-compatible format before the GPU rasterization exploit vector processing locally on the CPU.

Web Audio API provides audio synthesizing for Web Applications.
Web Audio applications create a graph of audio processing nodes that define the overall rendering behavior.
\libname{Webaudio} modules of Chromium and WebRTC similarly operate on an audio frame (2 channels of 128 audio samples) to:
(1) implement an audio node's functionality (GainNode that changes audio volume),
(2) transfer audio samples between audio node buffers,
or (3) merge multiple audio inputs.
\libname{PFFFT} is a Pretty Fast FFT implementation that provides frequency-domain analysis APIs for \libname{Webaudio}. 
Due to the latency constraints of Web Audio APIs, these libraries are not suited for domain-specific acceleration.

\libname{zlib} provides compression APIs for different libraries and applications, including \libname{libpng}.
\libname{zlib} uses two checksum algorithms (Adler-32 and CRC-32) to detect data corruption, which takes considerable execution time.
While both stages of compression (LZ77 and Huffman de/encode) are scalar, \libname{zlib} uses vector processing for checksum calculations.

\libname{boringssl} is Google's fork of OpenSSL, an SSL library that provides both low-level cryptography primitives (AES, DES, ChaCha20, and SHA256) and SSL implementation for secure network communications.
\libname{Arm Optimized Routines} contains an optimized version of math, network, and string utilities for Arm ISA-based processors.
We target string routines (memcmp, memchr, memcpy, and strlen) as many mobile applications extensively use them.

Due to the fine-grain \libname{zlib}, \libname{boringssl}, and \libname{Arm Optimized Routines} utility APIs that are extensively interleaved with applications' scalar code, domain-specific accelerations are not suited.

Opus is an audio codec for interactive audio applications.
\libname{libopus} is an Opus coder whose kernels operate on an audio frame.
We benchmark multiple kernels that:
(a) apply audio filters (Autoregressive Moving Average (ARMA) and Linear Predictive Coding (LPC) filters),
and (b) calculate autocorrelation (pitch and frequency autocorrelations).

\libname{libvpx} is the reference implementation of VP8 and VP9 video codecs, widely used by content providers like YouTube and Netflix.
We benchmark kernels that are common among most video codecs.
These kernels operate on an block of pixels to:
(a) calculate frequency transforms (forward and inverse Discrete Cosine Transform),
(b) compute block similarities (\textit{e.g.}, Sum of Absolute Difference),
and (c) quantize pixels for a higher compression rate.

\libname{XNNPACK} provides optimized Neural Network primitives that are employed in the back-end of Machine Learning frameworks such as TensorFlow Lite and PyTorch.
We evaluate four precisions for General Matrix Multiply (GEMM) and Sparse-Dense Matrix Multiplication (SpMM) kernels.
Various neural network APIs (such as convolutional and fully-connected layers) use these two kernels.
\section{Methodology and Tools}~\label{sec:methodology}

\vspace{-5mm}

\subsection{Workloads and Input Data Size}

\swan{} benchmark suite generates random inputs for workloads with the following sizes:
(a) 720x1280 (HD) images for image processing, graphics, and video processing libraries,
(b) 1 second of a standard audio stream with a 44.1 kHz sample rate for audio processing libraries,
(c) 128 KB data for data compression, cryptography, and string utility libraries,
and (d) 156 layers of Convolutional Neural Networks for the machine learning library.
When input data values affect the control flow of the workload, we carefully set the configuration and generate inputs based on the source library configurations and testing infrastructures.
To amortize the measurement errors, we repeat the measurement for multiple \textit{iterations} until the execution time reaches 1 second.
Finally, we ensure the correctness of \swan{} workloads by comparing the Scalar and Arm Neon implementation outputs.

\subsection{Evaluation Environment}

We analyze \swan{} workloads on Snapdragon 855 SoC, which enjoys Arm Big.LITTLE architecture with three different CPU configurations:
1 Prime and 3 Gold high-performance Cortex-A76 cores, and 4 Silver efficient Cortex-A55 cores.
We pin the \swan{} benchmark process in all experiments to the high-performance Prime core.
In Section~\ref{sec:perf-vs-core}, we show the sensitivity of vector processing performance to the core microarchitecture.
To focus on vector processing performance, we study single-thread implementation to minimize the error due to the multi-threading overhead.

\begin{table}[t]
  \scriptsize
  \centering
  \caption{Qualcomm Snapdragon 855 - Cortex-A76 Prime Core Baseline}
  \vspace{-2mm}
  \begin{tabular}{|c|c|}
    \hline 
    \textbf{Configuration} & \textbf{Detail} \\
    \hline \hline
    \multirow{2}{*}{Scalar core} & 2.8GHz, 128 entry ROB, out-of-order \\
    & 4-way Decode, 8-way Issue, 4-way Commit \\
    \hline
    Vector engine & 2 128-bit Advance SIMD units + crypto and FP16 ext \\
    \hline
    L1-I cache & 64 KiB, 4-way, 4 cycle latency  \\
    \hline
    L1-D cache & 64 KiB, 4-way, 4 cycle latency  \\
    \hline
    L2 cache & 512 KiB, 8-way, Private, Inclusive, 9 cycle latency \\
    \hline
    LLC & 2 MiB, 8-way, Shared, Inclusive, 31 cycle latency  \\
    \hline
   \end{tabular}
  \label{tab:config}
 \vspace{-3mm}
\end{table}

\subsection{Measurement and Simulation Tools}

We employ \textit{Android NDK r23c} to cross-compile \swan{} with \textit{Android Clang 12.0.9}.
We choose level 3 optimization for all workloads and compile the scalar code with auto-vectorization (\textit{Auto} implementation) to study its benefits for our set of data-parallel kernels.
We disable auto-vectorization for the scalar code in the rest of the evaluation (\textit{Scalar} implementation).
In addition, Clang replaces pieces of code with certain standard C library functions (such as \textit{memset}) that use Arm Neon acceleration.
Therefore, we also disable optimizations of standard C library functions for the \textit{Scalar} implementation.
With these compilation flags, we ensure that the \textit{Scalar} implementation does not employ any vector operations.
We also compiled the explicitly-vectorized Neon code with auto-vectorization, yet, our evaluation showed negligible auto-vectorization improvement.
Therefore, we only study the Neon code without auto-vectorization (\textit{Neon} implementation).
In addition, we minimize memory stalls to focus on the vector processing benefits by warming up caches before each \textit{iteration}.

We measure average battery current and voltage while executing each benchmark to calculate the chip's power consumption, including main memory.
Energy consumption is calculated as a product of power consumption and execution time.
We use \textit{Simpleperf}~\cite{simpleperf} to profile Performance Monitor Unit (PMU) events~\cite{pmuevents} of the device for all workloads to study microarchitectural bottlenecks in Section~\ref{sec:bottleneck}.





To study the scalability of vector processing with wider registers, we implement eight kernels with fake wider Neon intrinsics.
We develop a functional simulator for Neon intrinsics to ensure the correctness of the fake intrinsics.
Next, we capture dynamic instruction traces using a \textit{DynamoRIO}~\cite{dynamorio} client running on a server-class CPU with Armv8.2-A architecture (same as the mobile processor).
Finally, we develop a trace-driven simulator based on Ramulator~\cite{ramulator} CPU model and simulate instruction traces.
\section{Performance Analysis}\label{sec:perfenergy}


We analyze the instruction set, performance, and energy improvement of vectorized \swan{} workloads. Due to the limited of space, we show the Geometric mean of performance and energy improvements for kernels of each library.

\subsection{Instruction Diversity Analysis}

\begin{figure}[h]
    \centering
    \vspace{-3mm}
    \includegraphics[width=0.48\textwidth]{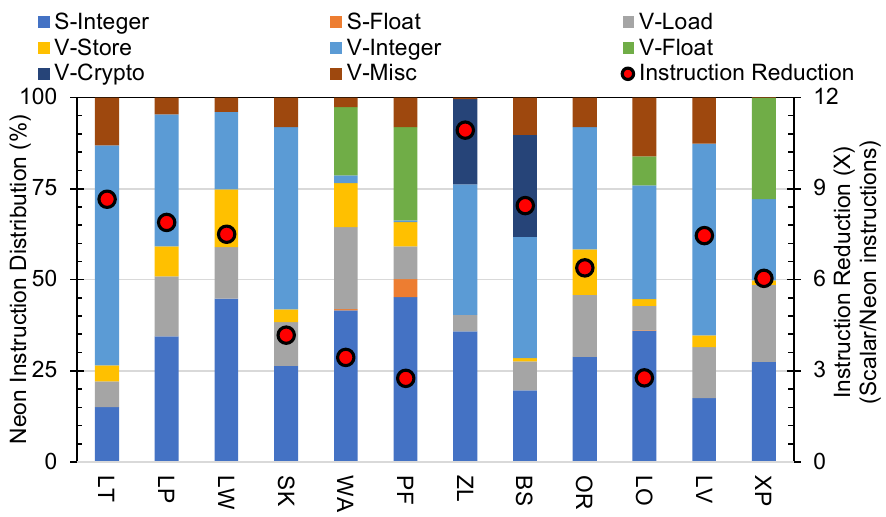}
    \vspace{-7mm}
    \caption{\underline{S}calar and \underline{V}ector instruction breakdown (primary Y-axis) of Arm Neon kernels and total instruction reduction (secondary Y-axis) compared to the Scalar implementation.}
    \label{fig:instr_breakdown}
\end{figure}

Figure~\ref{fig:instr_breakdown} (primary Y-axis) shows the instruction distribution of the Arm Neon implementation based on Arm Software Optimization Guide~\cite{cortexa76swopt}.
\textit{S-Integer + S-Float} shows the scalar part of the kernels.
\libname{PF} requires significant scalar computation for FFT calculation.
Hence, this library contains the highest portion of scalar instructions.
\libname{LT} linearly operates on an image row that requires the smallest portion of scalar instructions for control flow and address calculation.
A higher fraction of \textit{Vector Load and Store} instructions shows memory-intensive libraries with lower compute density.
\libname{WA} takes advantage of \textit{Vector APIs} (Section~\ref{sec:vect_library}) that keeps vector elements in memory, requiring a load and a store for each vector operation.
\swan{} contains libraries with different data types (integer and floating-point).
Audio processing and machine learning libraries (\libname{WA}, \libname{PF}, \libname{LO}, and \libname{XP}) contain floating-point operations.
\libname{XP} contains FP16 implementation of GEMM and SpMM.
\swan{} also provides a set of libraries that exploit cryptography acceleration of Arm Neon architecture, \textit{i.e.}, \libname{ZL} and \libname{BS}.
Finally, Vector Miscellaneous instructions are used for vector manipulation, \textit{i.e.}, converting vector register width and vector element type and combining/extracting multiple vector registers.
\libname{LO} contains operations with different data types and extensively uses Vector Miscellaneous instructions to manipulate vector registers.
In summary, \textit{\swan{} workloads encompass various applications with diverse instruction sets.}

Figure~\ref{fig:instr_breakdown} (secondary Y-axis) compares the dynamic instruction count of the Scalar and Neon implementations.
Assuming a similar code structure, the maximum instruction reduction of vector processing is bounded by the number of Vector Register Elements (\textit{VRE}), calculated by Equation~\ref{eq:max_instred}.


\begin{equation} \label{eq:max_instred}
    VRE = \frac{Vector \ Register \ Width \ (128~bits)}{Element \ Data \ Width}
\end{equation}

This equation shows that instruction reduction improves with lower precision data types.
Image and video processing libraries (\libname{LT}, \libname{LP}, \libname{LW}, and \libname{LV}) process 8-bit pixel values, encoding more operations in a vector instruction.
\libname{ZL} and \libname{BS} exploit cryptography instructions of Arm Neon architecture that reduces dynamic instructions substantially.
Therefore, \textit{Vector ISA encodes low-precision operations more efficiently compared to the scalar instruction set.}



\subsection{Performance Analysis}~\label{sec:performanceanalysis}


\begin{figure}[t]
    \centering
    \includegraphics[width=0.48\textwidth]{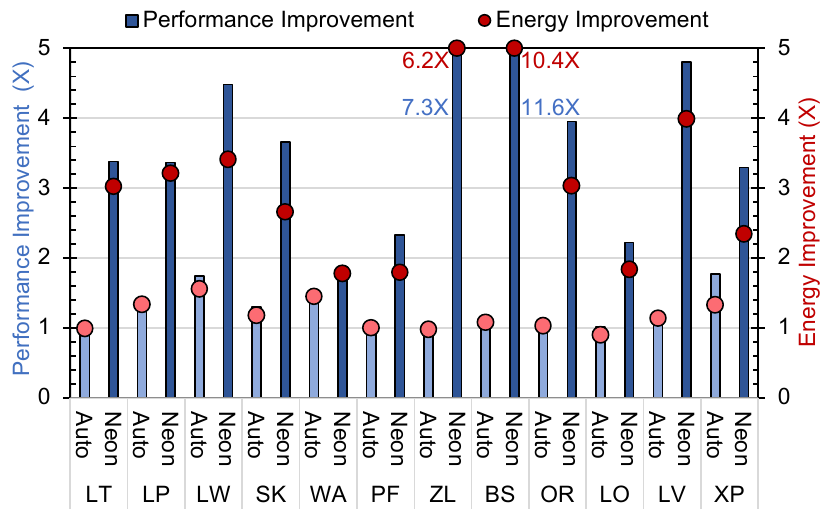}
    \vspace{-3mm}
    \caption{Performance (primary Y-axis) and Energy (secondary Y-axis) improvement of Auto-vectorization (Auto) and Explicit vectorization using Arm Neon intrinsics (Neon) compared to the Scalar implementation.}
    \vspace{-5mm}
    \label{fig:performance_energy}
\end{figure}

Figure~\ref{fig:performance_energy} (primary Y-axis) shows the performance of auto-vectorization (\textit{Auto}) and vector implementations (\textit{Neon}) normalized to the \textit{Scalar} implementation.
\textit{Vector processing, theoretically, improves the performance of the data-parallel portion of the program by \textit{VRE}}.
To explain this claim, we break the instructions of the data-parallel portion of the kernels into three categories:
(a) Address calculation and control flow scalar instructions:
Data-parallel kernels only need to calculate the base address of each vector memory access, which reduces the address calculations by \textit{VRE}.
Moreover, loop trip counts and their required control-flow instructions drop by a factor of \textit{VRE}.
(b) Efficient vector loads and stores:
A 128-bit vector load/store accesses \textit{VRE} data elements from the memory hierarchy.
The scalar kernel requires $VRE\times$ more cache accesses to load/store the same amount of data.
Assuming a high cache hit rate for data-parallel applications, vector loads and stores take $VRE\times$ fewer cycles.
(c) Vector compute instructions encode \textit{VRE} operations.
Assuming the same number of scalar and vector functional units with the same compute throughput in the microarchitecture, vector compute instructions provide a speedup factor of \textit{VRE}.

\textbf{Neon Performance Improvement.}
\libname{ZL} and \libname{BS} provide higher speedup because these libraries exploit the cryptography acceleration of Arm Neon architecture.
Except these libraries, Neon speedup is between $1.9\times$ (\libname{WA}) to $4.8\times$ (\libname{LV}).

According to Amdahl's law, Neon speedup decreases in kernels with a more significant portion of scalar code.
\libname{LO} requires pre-processing input and post-processing output data of the data-parallel portion of the code.
\libname{PF} frequently calculates the address and value of the FFT coefficients for the data-parallel computation.
The Neon implementations of \libname{LO} and \libname{PF} libraries achieve low dynamic instruction reduction (both $2.7\times$) and speedup ($2.2\times$ and $2.3\times$).

Our discussion showcases that total speedup correlates with \textit{VRE}.
\libname{XP} provides a speedup of $3.3\times$.
While the theoretical maximum speedup of 32-bit and 16-bit kernels is $4\times$ and $8\times$, our evaluation shows that the performance improvement of FP32/INT32 and FP16/INT16 kernels are in the range of [$2.0\times$ to $3.0\times$] and [$3.5\times$ to $5.1\times$], respectively.
In addition, audio processing libraries (\libname{WA}, \libname{PF}, and \libname{LO}) contain FP32 operations that drops \textit{Neon} speedup to $1.8\times$, $2.3\times$, and $2.2\times$, respectively.
\textit{Vector instructions encode low-precision operations more efficiently, increasing the speedup.}

Image processing libraries (\textit{e.g.}, \libname{LT} and \libname{LP}) require large working set sizes.
A large working set drops L1D, L2D, and LLC cache hit rates to 91\%, 90\%, and 67\% for the \libname{LT} library.
Therefore, memory accesses take more cycles and vector load latency is close to scalar load latency.
In this case, the speedup of vector memory accesses is lower than \textit{VRE}.
Hence, although \textit{LT} and \textit{LP} contain low precision operations on 8-bit pixel data and provide high \textit{VRE}, the achieved speedup of the \textit{Neon} implementation of both libraries is $3.3\times$.
\textit{Lower cache hit-rate drops vector processing speedup in data-parallel kernels with large working set size.}

\textbf{Auto-Vectorization Performance Improvement.}
It is believed that auto-vectorization requires less effort from programmers than explicit vectorization.
However, our analysis shows that \textit{auto-vectorizing the legacy scalar code that is optimized for higher super-scalar performance is not achieving sufficient speedups}.
In fact, auto-vectorization requires complex loop transformations~\cite{vectortransformations} to expose the DLP to the compiler.
Table~\ref{tab:auto_perf} shows that among the 59 analyzed data-parallel kernels, auto-vectorization enhances the performance of only 23 kernels.
From these 23 kernels, \textit{Auto} marginally outperforms 5 \textit{Neon} implementations only because of higher loop interleaving count.

\begingroup
\setlength{\tabcolsep}{3pt} 
\renewcommand{\arraystretch}{1} 
\begin{table}[ht]
\scriptsize
  \renewcommand\arraystretch{1}
  \centering
  \caption{\textit{Auto} Performance W.R.T \textit{Scalar} and \textit{Neon}.}
  \begin{tabular}{|c|c||c|c|}
    \hline
    \textbf{\textit{Auto} vs. \textit{Scalar}} & \textbf{\#Kernels} & \textbf{\textit{Auto} vs. \textit{Neon}} & \textbf{\#Boosted Kernels} \\ \hline
    $Auto \approx Scalar$ & $34$ & $Auto \approx Neon$ & 6 \\ \hline
    $Auto < Scalar$ & $2$ & $Auto < Neon$ & $12$ \\ \hline
    $Auto > Scalar$ & $23$ & $Auto > Neon$ & $5$ \\ \hline
   \end{tabular}
  \label{tab:auto_perf}
\end{table}
\endgroup


In general, our study shows two main reasons for auto-vectorization failure:

\textbf{First}, \textit{compiler is unable to prove the legality of vectorization, \textit{i.e.}, the safety and correctness of vector transformations.}
We provide three examples that fail the legality of vectorization and how \textit{Neon} implementations solve them.

\underline{Example 1.}
Loops shall be countable, \textit{i.e.}, compiler can calculate the number of loop iterations based on the variables.
A \texttt{for} loop with a \texttt{break} statement or a \texttt{while} loop with an unknown condition is not countable.
\textit{Uncountable loops prevent vectorization in eight data-parallel kernels.}
Neon uses \texttt{reduction} instructions to detect loop break conditions.

\underline{Example 2.}
Compilers use run-time checks if there is not enough information in compile-time to prove the safety of the vectorization.
Code patterns that hinder run-time checks fail vectorization.
For example, indirect memory accesses such as \texttt{A[B[i]]} prevent \textit{memory aliasing} checks as calculating the boundary of accesses to array \texttt{A} requires evaluating all elements of array \texttt{B}.
This code pattern is used to convert computations to look-up tables and optimize the scalar code for super-scalar performance.
\textit{Indirect memory accesses fail compiler vectorization in 8 data-parallel kernels.}

\underline{Example 3.}
Variables are often initialized before the loops and used and modified inside the loop.
In this case, compilers add a \texttt{PHI} node that selects between the initial values (before the loop) and modified values (within the loop).
Therefore, \texttt{PHI} nodes generate data dependencies between different iterations of the loop that the vectorizer must appropriately handle.
LLVM can recognize and handle the common \texttt{PHI} nodes in vectorization.
However, complex \texttt{PHI} nodes fail compiler vectorization because of data dependencies.
For example, \textit{Downsample} kernel of \libname{LT} initializes bias values before the loop
and used and modifies biases inside the loop.
LLVM auto-vectorizer fails as it cannot safely resolve this data dependency.
Neon implementation, however, uses pre-defined constant bias values.
\textit{Loop data dependencies prevent compiler vectorization in 9 kernels of \swan{} benchmark suite.}

Other legality obstacles, such as \textit{reordering Floating-Point and Memory operations}, \textit{inability to vectorize \texttt{CALL} instructions and \texttt{switch} statements}, and \textit{unsafe memory operations} prevent compiler vectorization in 10 kernels.

\textbf{Second}, compiler employs a heuristic cost model that compares the benefits of vectorizing a loop with different Vectorization Factors ($VF$).
Then, it chooses the $VF$ with the minimum cost-to-width ratio.
LLVM cost model, however, suffers from inaccuracies because different code characteristics (\textit{e.g.,} loop trip counts and control flow behavior) and microarchitectural features (\textit{e.g.,} throughput and latency of each instruction) are not known in the compile time.
\textit{Inaccurate cost model prevents compiler vectorization in 12 kernels of the \swan{} benchmark suite.}

\subsection{Power and Energy Analysis}\label{sec:energyanalysis}

\begin{figure}[ht]
    \centering
    \includegraphics[width=0.48\textwidth]{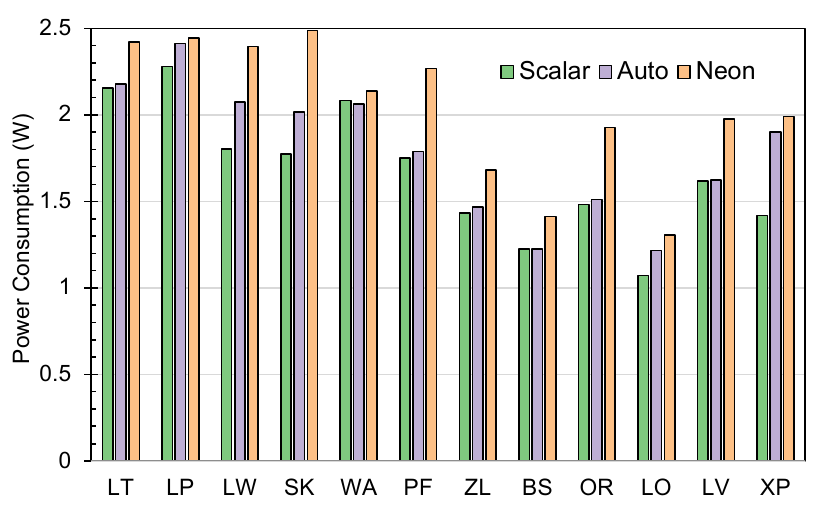}
    \vspace{-6mm}
    \caption{Total chip Power Consumption of libraries (including DRAM power).}
    \label{fig:power}
\vspace{-3mm}
\end{figure}

While vector processing substantially improves execution time, Figure~\ref{fig:power} shows that it increases total chip power consumption, \textit{including the main memory.}
This is because
vector processing increases DRAM access rate.
Since memory accesses consume significant power, an increased DRAM access rate results in higher power consumption.

We measure the number of main memory accesses using LLC misses and the main memory access rate as the number of memory accesses per cycle.
Our evaluation shows that the \textit{Neon} memory access rate is $8.8\times$ more than \textit{Scalar} implementation.
This is because of the higher LLC miss rate and lower cycles of the \textit{Neon} implementation.
Comparing the power consumption of libraries shows that kernels with higher LLC miss rate and memory access rate, such as the image processing and graphics libraries (\libname{LT}, \libname{LP}, \libname{LW}, and \libname{SK}), consume more chip power.

Figure~\ref{fig:performance_energy} (secondary Y-axis) shows the energy consumption of \textit{Auto} and \textit{Neon} normalized to the \textit{Scalar}.
Despite the higher power consumption of \textit{Neon} implementations, significantly lower execution time and dynamic instruction count result in high energy savings, specifically in kernels with lower precision (higher \textit{VRE}) and higher cache hit rate, such as \libname{LW} ($4.4\times$), \libname{OR} ($3.9\times$), and \libname{LV} ($4.8\times$).

\subsection{Bottleneck Analysis}\label{sec:bottleneck}

To study the bottlenecks of vector processing, we follow Intel's top-down microarchitecture analysis~\cite{topdown} by profiling microarchitectural characteristics of data-parallel kernels.
Table~\ref{tab:pmu} studies cache (L1D, L2, and LLC) Miss Per Kilo Instructions (MPKI), the portion of cycles stalled by Front-End and Back-End (\%), and Instructions Per Cycle (IPC).

Data-parallel kernels enjoy regular control flow.
Therefore, front-end stalls are less than 5\% in the Neon implementations.
In 4 data-parallel kernels, control divergence is handled by \textit{If-Conversion}, where control dependency is converted to data dependency by executing both branches and choosing the final values of a vector using \texttt{AND/OR} or \texttt{BSL} (Bitwise Select) instructions.
Low front-end stalls show that the \textit{front-end modules of the microarchitecture, such as Instruction Cache and TLB, Branch Predictor, and Instruction Fetch and Decode 
are not a bottleneck for vector processing}.

IPC is lower in Neon implementations due to higher Back-End stalls of data-parallel kernels.
We observe that two ASIMD functional units of the Prime core provide enough computation throughput.
Therefore, \textit{data-parallel applications with high back-end stalls are mainly bounded by memory stalls.}
Vector processing increases the pressure on the cache hierarchy by frequent cache accesses.
Hence, all cache levels experience higher MPKI.
Scalar and Neon implementations of \libname{XP} manually unroll the loops with a factor of 32 and 8.
Thus, a burst of memory accesses is injected into the L1D cache, increasing L1D MPKI of \textit{Scalar} and \textit{Neon} to $4.9$ and $20.1$.
Workloads with larger working set sizes, such as image processing and graphics libraries (\libname{LJ}, \libname{LP}, and \libname{SK}), increase L2 and LLC MPKI.

\begingroup
\setlength{\tabcolsep}{3pt} 
\renewcommand{\arraystretch}{1} 
\begin{table}[ht]
\scriptsize
  \renewcommand\arraystretch{1}
  \centering
  \caption{microarchitectural Characteristics of Evaluated Libraries.}
  \scalebox{1}{
  \begin{tabular}{|c||c|c||c|c||c|c||c|c||c|c||c|c|}
    \hline
    \multirow{2}{*}{\textbf{Lib-}} & \multicolumn{2}{c|}{\textbf{L1D}} & \multicolumn{2}{c|}{\textbf{L2}} & \multicolumn{2}{|c|}{\textbf{LLC}} & \multicolumn{2}{|c|}{\textbf{Front-End}} & \multicolumn{2}{|c|}{\textbf{Back-End}} & \multicolumn{2}{|c|}{\textbf{IPC}} \\
    \multirow{2}{*}{\textbf{rary}} & \multicolumn{2}{c|}{\textbf{MPKI}} & \multicolumn{2}{c|}{\textbf{MPKI}} & \multicolumn{2}{|c|}{\textbf{MPKI}} & \multicolumn{2}{|c|}{\textbf{Stalls (\%)}} & \multicolumn{2}{|c|}{\textbf{Stalls (\%)}} & \multicolumn{2}{|c|}{\textbf{}} \\
    \cline{2-13}
    & \textbf{S} & \textbf{V} & \textbf{S} & \textbf{V} & \textbf{S} & \textbf{V} & \textbf{S} & \textbf{V} & \textbf{S} & \textbf{V} & \textbf{S} & \textbf{V} \\
    \hline \hline
    \libname{LJ} & 1.7 & 17.0 & 0.5 & 7.4 & 2.8 & 27.0 & 0.2 & 0.2 & 14.9 & 51.9 & 3.04 & 1.2 \\ \hline
    \libname{LP} & 1.2 & 8.1 & 0.3 & 6.5 & 2.8 & 17.9 & 0.2 & 0.4 & 11.3 & 47.2 & 2.9 & 1.4 \\ \hline
    \libname{LW} & 0.2 & 2.4 & 0.1 & 0.0 & 0.1 & 0.1 & 0.1 & 0.1 & 11.4 & 38.4 & 2.1 & 1.2 \\ \hline
    \libname{SK} & 1.2 & 4.8 & 0.1 & 0.4 & 2.3 & 9.4 & 0.1 & 0.1 & 10.4 & 25.2 & 2.6 & 2.3 \\ \hline
    \libname{WA} & 0.1 & 0.2 & 0.1 & 0.1 & 0.1 & 0.1 & 0.2 & 3.2 & 1.1 & 10.3 & 3.1 & 2.7 \\ \hline
    \libname{PF} & 3.1 & 9.7 & 1.2 & 3.3 & 0.1 & 0.1 & 0.4 & 0.2 & 22.3 & 28.2 & 2.9 & 2.4 \\ \hline
    \libname{ZL} & 0.8 & 5.8 & 0.1 & 0.1 & 0.1 & 0.1 & 0.1 & 0.4 & 12.1 & 25.9 & 3.3 & 2.1 \\ \hline
    \libname{BS} & 0.5 & 1.2 & 0.1 & 0.1 & 0.1 & 0.1 & 0.1 & 0.1 & 11.0 & 34.5 & 2.6 & 2.2 \\ \hline
    \libname{OR} & 1.2 & 6.2 & 0.1 & 0.1 & 0.1 & 0.1 & 0.2 & 0.2 & 4.3 & 27.3 & 3.3 & 2.2 \\ \hline
    \libname{LO} & 0.8 & 2.4 & 0.1 & 0.2 & 0.1 & 0.1 & 0.4 & 0.1 & 27.3 & 49.7 & 2.1 & 1.6 \\ \hline
    \libname{LV} & 1.4 & 7.8 & 0.3 & 3.3 & 1.9 & 6.5 & 0.2 & 0.1 & 15.9 & 41.9 & 2.8 & 1.6 \\ \hline
    \libname{XP} & 4.9 & 20.0 & 0.5 & 1.3 & 0.6 & 2.0 & 0.2 & 0.2 & 35.7 & 46.3 & 2.1 & 1.5 \\ \hline
    \multicolumn{13}{l}{*Numbers are rounded up to 1 decimal place.} \\
    \multicolumn{13}{l}{*S and V are \underline{S}calar and \underline{V}ectorized Neon implementations.}
   \end{tabular}
   }
  \label{tab:pmu}
\end{table}
\endgroup


\subsection{Performance vs. Core Architecture}~\label{sec:perf-vs-core}

\vspace{-3mm}

\begin{figure}[h]
    \centering
    \vspace{-3mm}
    \includegraphics[width=0.48\textwidth]{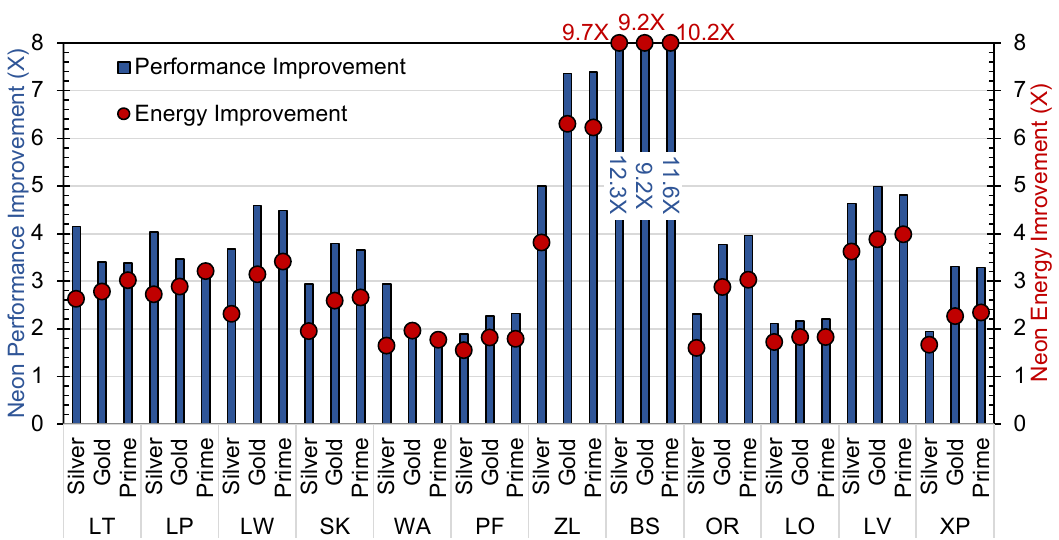}
    \vspace{-6mm}
    \caption{Performance analysis of core architectures.}
    \label{fig:cores}
\end{figure}

To further study the sensitivity of vector processing to core microarchitecture, Figure~\ref{fig:cores} shows the performance (Primary Y-axis) and energy (Secondary Y-axis) improvement of Neon normalized to the Scalar execution.
Silver (Cortex-A55) core contains an In-Order pipeline with one 128-bit ASIMD functional unit at 1.8 GHz, while Gold and Prime (Cortex-A76) enjoy two ASIMD functional units of an out-of-order microarchitecture at 2.4 and 2.8 GHz, respectively.

Comparing Silver with Gold and Prime cores shows us that more ASIMD units do not substantially improve performance and energy consumption.
In fact, more ASIMD units only benefit when there is enough Instruction-Level Parallelism (ILP).
When a vector compute instruction is data dependent on a vector load/store, it is not issued to the ASIMD units until the memory access is finished and the vector load/store is executed.
Thus, higher memory stalls of \libname{LT}, \libname{LP}, and \libname{LO} libraries reduce ILP and Neon performance and energy improvement.
\libname{XP} implementations manually unroll the loops; therefore, this library increases the ILP, and Gold and Prime cores enjoy more ASIMD functional units and out-of-order processing.

Silver cores require less power consumption due to the in-order pipeline and lower DRAM access rate.
Silver and Gold cores consume less power than the Prime core due to lower frequencies.
Consequently, Neon achieves higher energy savings for Prime cores in nearly all workloads.

\section{Common Computation Patterns}~\label{sec:patterns}


In this section, we study common code patterns of the evaluated libraries and discuss their bottlenecks.

\subsection{Reduction}

Reduction is a computation pattern that reduces vector elements into a single result.
\swan{} benchmark suite contains seven data-parallel kernels with reduction operations.
These kernels exploit parallelism for vectorization in two ways:

(1) \textbf{Inter-Reduction Parallelism:}
\textit{Neon} exploits the parallelism \textit{between} multiple reduction operations.
Each vector instruction performs 1 step of reduction for \textit{VRE} output results.
For example, \textit{Vertical/Horizontal Convolution} kernels of \libname{SK} convolve multiple columns/rows in parallel.
In this way, vectorization imposes no overhead on the kernel.

(2) \textbf{Intra-Reduction Parallelism:} Five kernels compute a single output in each invocation.
\textit{Neon} exploits the parallelism \textit{within} a single reduction operation.
\textit{When reduction function is both Associative and Commutative}, vector implementation breaks the reduction to \textit{VRE} partial reductions and processes \textit{VRE} inputs in each iteration.
The only overhead of this reduction pattern is reducing \textit{VRE} partial results to one final result.
An example is the \textit{Audible} kernel of \libname{WA}, which calculates the energy of an audio frame as $\sum_{i=0}^{N}s_i^2$.

Vector implementation parallelizes the sequential reduction and changes the order of the operations AND source operands.
Consequently, \textit{a reduction operation must be Associative AND Commutative to be vectorized.}
Otherwise, the scalar reduction requires significant algorithm modifications to be vectorizable.


For example, \textit{Adler-32} kernel of \libname{ZL} calculates two checksum values for an array of $N$ characters: $S1 = \sum_{i=0}^{N}b_i$ and $S2 = \sum_{i=0}^{N}(N-i)\times b_i$.
Scalar implementation sequentially consumes characters and calculates $S1+=b_i$ and $S2 += S1$.
While the first reduction operation ($S1$) is associative and commutative, the second ($S2$) is neither associative nor commutative.
A na\"ive approach to vectorize these reduction patterns is loop distribution.
For example, we can calculate $S1$ in a separate loop and store partial values ($PS1_i$) in the memory.
In the second loop, we modify $S2$ reduction operation to $S2 = \sum_{i=0}^{N}PS1_i$.
This way, both $S1$ and $S2$ reduction operations are associative, commutative, and, hence, parallelizable.
\swan{} benchmark suite contains five data-parallel kernels with this computation pattern.

\subsection{Random Memory Access}

Gather and Scatter operations access arbitrary memory locations, enabling random access patterns in 7 data-parallel kernels of \swan{} benchmark suite.
While RISC-V Vector Extension (RVV)~\cite{riscv} implements random memory accesses (a.k.a, \textit{Indexed Vector Load/Store} intrinsics), Arm Neon lacks a general-purpose solution for these access patterns.

We observe that random access patterns in all seven data-parallel kernels are employed for gathering values from \textit{Look-Up Tables} for two reasons:
(a) transforming vector elements (keys) to another domain (values).
or (b) converting multiple operations to single look-up table access.
For example, in each round of \textit{AES} cipher, \libname{BS} substitutes AES states (keys) with other states (values).
The following code listing shows the semantics of look-up table accesses in these kernels.

\begin{lstlisting}[style=CStyle]
template <typename T1, typename T2>
void LU_TBL(T1*table, T2* keys, T1* vals, int len)
    for (int i = 0; i < len; i++)
        vals[i] = table[keys[i]];
\end{lstlisting}

If the table contains less than 64 8-bit values, one can load it in multiple vector registers and access the table registers using Arm Neon's \texttt{TBL} instructions.
This pattern is not employed in these data-parallel kernels since the table contains more than 64 values.
Arm Neon provides domain-specific acceleration for two kernels with Look-Up Table access, \textit{i.e.}, \textit{AES} and \textit{CRC-32} Cryptography intrinsics.
Developing Look-Up Table access for \textit{Neon} requires exporting elements of \texttt{key} vector register to the Scalar registers, performing Look-Up Table access for each individual key using scalar instructions, and packing the results back to the \texttt{value} vector register.
Due to the high overhead of this operation, four kernels give up the benefits of look-up tables, and one kernel (\textit{DES}) of \libname{BS} does not provide \textit{Neon} implementation.

While we exclude \textit{DES} kernel from this paper's evaluation, we developed a \textit{Neon} implementation to study the overhead of Look-Up Table accesses in vectorization without Random Memory Access intrinsics.
Our evaluation shows $11\%$ slow-down compared to \textit{DES} scalar implementation.
Next, we deprecated Look-Up Table accesses from the \textit{Scalar} and \textit{Neon} baselines.
In this case, \textit{Neon} outperforms \textit{Scalar} by $2.1\times$.
Our evaluation shows Table Look-Up accesses take $73\%$ of total instructions in \textit{DES} Neon implementation.

Therefore, \textit{supporting intrinsics for gathering values from Look-Up Tables benefits seven data-parallel workloads.}


\subsection{Strided Memory Access}

Arm Neon supports memory accesses with stride values up to 4.
4-stride memory accesses are frequently used in Image Processing and Graphics libraries to \textit{load and de-interleave} or \textit{interleave and store} 4-channel pixel values.
Non-unit stride memory accesses are implemented using Arm's multi-register data types and instructions.
A multi-register data type, \texttt{TWxExR\_t} encompasses \texttt{R} vector registers, each of which contains \texttt{E} elements (\textit{i.e.}, \textit{VRE}) of \texttt{W}-bit \texttt{T}-type data.
The following code listing shows a strided memory load that fetches $E(16) \times R(4)$ data values and interleaves them between $R(4)$ vector registers.
Therefore, adjacent vector elements within a vector register are loaded from memory locations with stride $R = 4$.
A stride memory store operates in the opposite direction, where $E$ elements from $R$ registers are stored in memory with the stride of $R$.

\begin{lstlisting}[style=CStyle]
uint8x16x4_t vld4q_u8(uint8_t const *ptr) {
    uint8x16x4_t Result;
    for (int element = 0; element < 16; element++)
        for (int reg = 0; reg < 4; reg++)
            Results[reg][element] = *ptr++;
    return Result;
}
\end{lstlisting}

Arm Neon also supports \texttt{UZP} (\textit{de-interleave}) and \texttt{ZIP} (\textit{interleave}) instructions, which, instead of accessing memory, move elements between vector registers with the stride value of 2.
Table~\ref{tab:strided_ldst} shows the number of kernels containing non-unit stride value instructions and the average portion of these instructions in those kernels.

\begingroup
\setlength{\tabcolsep}{3pt} 
\renewcommand{\arraystretch}{1} 
\begin{table}[h]
\scriptsize
  \renewcommand\arraystretch{1}
  \centering
  \caption{Number of Kernels and Portion of Strided Memory Accesses.}
  \begin{tabular}{|c|c|c|c|}
    \hline
    \textbf{Stride} & \textbf{Instruction} & \textbf{\#Kernels} & \textbf{Avg. Portion} \\
    \hline
    \multirow{4}{*}{2} & \texttt{LD} & 1 & $2.9\%$ \\
    \cline{2-4}
    & \texttt{ST} & 4 & $2.3\%$ \\
    \cline{2-4}
    & \texttt{ZIP} & 5 & $6.2\%$ \\
    \cline{2-4}
    & \texttt{UZP} & 7 & $3.0\%$ \\
    \hline
    \multirow{2}{*}{4} & \texttt{LD} & 8 & $5.8\%$ \\
    \cline{2-4}
    & \texttt{ST} & 8 & $4.7\%$ \\
    \hline
    \end{tabular}
  \label{tab:strided_ldst}
\end{table}
\endgroup

\textit{Arm Neon multi-register data types and instructions efficiently encode memory accesses with non-unit strides of up to 4 in Image Processing and Graphics libraries.}
However, when a higher stride value is required, one needs to use multiple instructions that hurt the performance of the kernel.
\textit{RVV} architecture implements \textit{Strided Vector Load/Store}, which can efficiently encode arbitrary strides memory accesses.

\subsection{Matrix Transposition}

Matrix Transposition is a common computation pattern used in six data-parallel kernels.
In addition, Matrix Transposition is one of the frequently-used primitives of \libname{XP} to transpose the input of Neural Network layers for cache-friendly memory access.
\textit{FFT} kernels of \libname{PF} use eight \texttt{ZIP} instructions to transpose a $4\times4$ matrix of 32-bit floating-point numbers.
Matrix transposition is only used in pre- and post-processing steps of FFT using on average $3.3\%$ of the total \libname{PF} instructions.
Arm Neon provides transpose intrinsics that use two instructions to transpose every two elements of two vector registers using the following code listing.

\begin{lstlisting}[style=CStyle]
int16x8x2_t vtrnq_s16(int16x8_t a, int16x8_t b) {
    int16x8x2_t Result;
    for (int element = 0; element < 8; element+=2) {
        // {a[0],b[0],a[2],b[2],...}
        Results[0][element] = a[element];
        Results[0][element+1] = b[element];
        // {a[1],b[1],a[3],b[3],...}
        Results[1][element] = a[element+1];
        Results[1][element+1] = b[element+1];
    }
    return Result;
}
\end{lstlisting}

\textit{Forward and Inverse DCT} kernels of \libname{LV} use intrinsics similar to the code listing above to transpose $8\times8$ \textit{DCT} blocks of 16-bit integer numbers in 32 instructions.
While this is the most efficient way of implementing $8\times8$ matrix transposition using Arm Neon, it takes, on average, $24.1\%$ of total \libname{LV} instructions.

To transpose an arbitrary-sized $M \times N$ matrix, one only needs to use $VRE \times VRE$ matrix transposition primitives to transpose each square sub-block separately.
Thus, we calculate the latency of matrix transposition ($lat_{A^T}$) using the following equation: \vspace{-2mm}

\begin{equation} \label{eq:transposition_cost}
\begin{aligned}
    & lat_{A^T}(M, N) = \lceil \frac{M}{VRE} \rceil \times \lceil \frac{N}{VRE} \rceil \times lat_{A^T}(VRE,VRE)
\end{aligned}
\end{equation}

For example, \libname{LV} library also requires matrix transposition of $16 \times 16$ DCT blocks of $16-bit$ values.
\libname{LV} breaks the block into four quarters and transposes them separately using the mentioned $8 \times 8$ matrix transposition primitive.
Therefore, \libname{LV} achieves $16 \times 16$ matrix transposition in $4 \times 32 = 128$ cycels.

\textit{Matrix Transposition is an intensive computation pattern of data-parallel kernels. \swan{} contains efficient matrix transposition primitives as a part of \libname{PF} and \libname{LV} libraries.}

\subsection{Portable Vector APIs}~\label{sec:vect_library}
While we showed that explicit vectorization achieves a higher speedup compared to auto-vectorization, performance improvement comes with the cost of re-developing kernels for a new ISA or ISA extension.
To avoid kernel re-development for a new ISA, applications add a new layer of abstraction using a library of \textit{Portable Vector APIs} with common simple functions and use these APIs across the data-parallel kernels.
Vector libraries provide a set of implementations for different ISAs, and select an implementation based on the capabilities of the target platform.
Therefore, to support a new ISA, applications are only required to implement vector APIs for the target instruction set, while the algorithm of the data-parallel functions remains the same.

\libname{PF} defines a set of macros for common vector intrinsics to operate on floating-point vector variables.
Due to the different instruction sets of various vector architectures, \libname{PF} only supports basic intrinsics and does not take advantage of sophisticated vector instructions.
For example, \libname{PF} uses a na\"ive vector \textit{Complex Multiplication} implementation that takes six instructions and eight cycles of Cortex-A76 core~\cite{cortexa76swopt}.
Armv8.2-A architecture supports multiply-add and multiply-subtract operations, with which Complex Multiplication requires four instructions and five cycles.
Armv8.3-A supports complex multiply-add intrinsics that only require one instruction and take two cycles of Arm Cortex-A710 core~\cite{cortexa710swopt}.
However, none of these sophisticated intrinsics are supported by Intel SSE; therefore, \libname{PF} only uses basic vector APIs, dropping \textit{Neon} performance improvement to only $2.3\times$.

\begin{figure*}[t]
    \centering
    \includegraphics[width=0.98\textwidth]{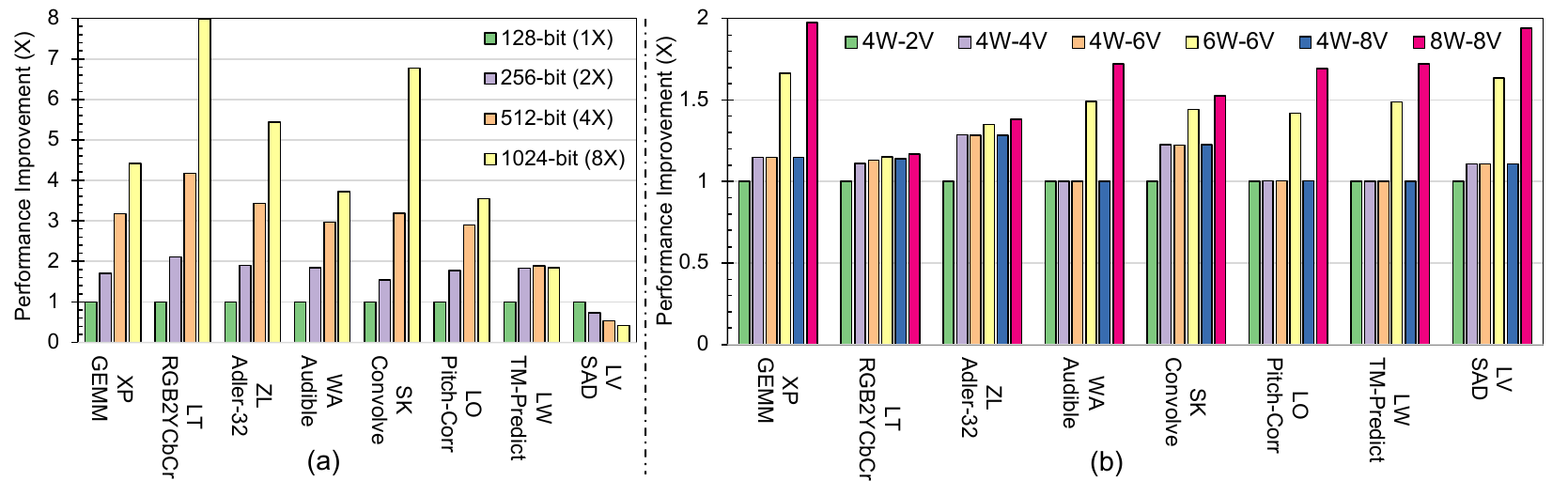}
    \vspace{-2mm}
    \caption{Neon performance scalability with: (a) wider vector registers and (b) more \underline{V}ector execution units and O.o.O \underline{W}ays.
    }
    \label{fig:scalability}
    \vspace{-3mm}
\end{figure*}

\textit{Webaudio} modules (\libname{WA}) of \textit{Chromium} and \textit{WebRTC} projects use vector APIs with simple vector operations such as vector convolution, multiplication, clip, etc.
These fine-grain vector APIs load input arrays in the vector variables, perform a simple operation, and store the results back to the output arrays.
Therefore, \libname{WA} requires a load and a store for every arithmetic operation.
In fact, around $59\%$ of \libname{WA}'s vector instructions are loads and stores, dropping instruction reduction to $3.4\times$ and \textit{Neon} speedup to $1.8\times$.

\textit{While Vector APIs substantially reduce the cost of supporting different vector processing architectures, they increase the number of instructions and significantly limit the benefits of vector processing.}
\section{Scalability Analysis}~\label{scalability}

\vspace{-3mm}

The evaluated Cortex-A76 Prime core baseline is equipped with two 128-bit ASIMD units.
We study the performance scalability with wider vector registers and more vector execution units using eight kernels with different computation patterns that are representative of their libraries.

\subsection{Wider Vector Registers}

We redevelop kernels with wider registers using a fake Arm Neon library with 128, 256, 512, and 1024-bit registers and further optimize the algorithms based on the available set of vector instructions for each vector register width.
Figure~\ref{fig:scalability}(a) shows the performance improvement of these implementations.
Note that 2/4/8$\times$ wider registers exploit data-level parallelism to improve the performance at the cost of 2/4/8$\times$ larger vector register file and ASIMD units.

Wider registers are beneficial when a data-parallel kernel only requires streaming memory accesses.
For example, \libname{LT}'s \textit{RGB-to-YCbCR} and \libname{SK}'s Convolution enjoy high SIMD lane utilization of $99\%$ and $98\%$.
Therefore, 1024-bit implementations improve performance by $7.9\times$ and $6.7\times$ compared to the 128-bit implementations using 8$\times$ SIMD lanes.

\textit{GEMM-FP32} implementation exploits parallelism across the output matrix columns.
When the number of output columns is not evenly divisible by \textit{VRE}, \textit{Neon} implementations use narrower registers that drop SIMD utilization from $98\%$ of 128-bit to $89\%$ of 1024-bit implementations.

\libname{WA}'s \textit{Audible} kernel measures the energy of an audio channel using reduction.
While Arm Neon efficiently reduces 128-bit vector registers to a scalar register with \texttt{{U/S}ADDLV} instructions, we do not extend these instructions to wider implementations.
This is because these instructions take 5/6 cycles for 128-bit vector registers~\cite{cortexa76swopt}, and reducing very wide registers results in longer latencies.
Instead, we reduce wider registers to 128-bit in multiple iterations by breaking them into two halves.
Hence, SIMD utilization and speedup of 1024-bit implementation drop to $74\%$ and $3.7\times$.

When kernels process small multi-dimensional input data, wider SIMD lanes require numerous vector manipulation instructions.
For example, \libname{LV}'s \textit{Sum of Absolute Differences (SAD)} and \libname{LW}'s \textit{TM-Prediction} kernels need to fetch data from $8\times 8$ and $16\times 16$ blocks of pixels.
While 128-bit implementation encodes fetching a row of the input data efficiently with one vector load instruction, wider registers require loading each row and packing them before processing.
These implementations are significantly bounded by the vector manipulation instructions due to the \textit{Neon}'s inability to encode multi-dimensional strided accesses efficiently.
Therefore, wider registers do not benefit these kernels.

\subsection{Increasing the Number of ASIMD Units}

Figure~\ref{fig:scalability}(b) shows the performance improvement of an Arm core with 2/4/6/8 128-bit ASIMD units (V).
These configurations take advantage of vector Instruction-Level Parallelism (ILP) at the cost of more vector register file ports and ASIMD units.
In addition, we employ 2/4/6/8 decode and commit ways (W) to issue enough instructions to the vector execution units.
4W-2V baseline is the evaluated Cortex-A76 core of Table~\ref{tab:config}.

We observe that increasing the number of vector execution pipelines to more than decode ways (\textit{i.e.}, 4W-6V and 4W-8V implementations) provides limited performance improvement.
This is because ASIMD units are under-utilized due to the lack of enough issued instructions.
Hence, the core is unable to exploit the ILP of the workloads.

In configurations with enough decode ways, we observe that ASIMD unit utilization is limited by the inherent ILP of the workload.
Note that we manually unroll loops to increase the vector ILP of the workloads.
\textit{GEMM} kernel of \libname{XP} simultaneously computes 32 output columns using eight 128-bit registers.
\libname{LV}'s \textit{SAD} kernel is unrolled to compute the Sum of Absolute Difference of a video frame in 32 accumulators of eight vector registers in parallel.
\textit{GEMM} and \textit{SAD} provide the highest vector ILPs.
Hence, 8W-8V configuration outperforms 4W-2V by 1.9$\times$ in both kernels using 4$\times$ ASIMD units.

\textit{RGB-to-YCbCr} kernel of \libname{LT} converts the color space of 16 pixels in four vector registers in parallel.
However, due to the high vector register pressure of this kernel, LLVM's bottom-up list instruction scheduler~\cite{llvmcodebook} schedules vector instructions close to their use and reduces the ILP.
Therefore, ILP is limited to computing three color spaces of four output pixels, limiting the 8W-8V configuration performance to 1.2$\times$ of the 4W-2V configuration (4$\times$ ASIMD units).


\vspace{-3mm}

\section{Mobile Application Processors}~\label{sec:appprocessor}

\vspace{-3mm}

Mobile SoCs are equipped with different domain-specific accelerators such as GPU and DSP.
These can be alternatives to vector processing for data-parallel kernels. 
However, domain-specific accelerators suffer from data transfer and kernel launch overheads negating their acceleration benefits for fine-grain kernels.
Vector processing, on the other hand, enjoys tight integration to the CPU and fine-grain instruction interleaving without data transfer or kernel launch overhead.

We observe that the first nine libraries of Table~\ref{tab:libraries} are not accelerated on GPUs.
Table~\ref{tab:gpu_dsp_overhead} compares \textit{only} kernel launch overhead with the average \textit{Neon} execution time of these libraries.
We measure the time of launching a dummy kernel on Adreno 640 GPU (Qualcomm OpenCL driver) and Hexagon 690 DSP (fastRPC).
On average, GPU and DSP kernel launch overhead alone is $1.9\times$ and $19\%$ of the average execution time of the aforementioned nine libraries on Neon.
These libraries provide fine-grained APIs that are necessary for workloads with parallel and serial code interleaving.
Therefore, GPU and DSP are not employed in any of these libraries due to the data transfer cost and kernel launch overhead.
In addition, GPU suffers from high power consumption, which reduces the battery life of mobile devices. Programming DSP is complex, and it only supports fixed-point operations.

\begin{table}[ht]
  \small
  \centering
  \scriptsize
  \vspace{-2mm}
  \caption{GPU and DSP kernel launch overhead vs. Neon total execution time.}
  \scalebox{0.95}{
  \begin{tabular}{|c|c||c|c|c|}
    \hline
    \multicolumn{2}{|c||}{{Kernel Launch Overhead}} & \multicolumn{3}{c|}{{Neon Kernel Execution}} \\
    \hline
    Adreno 640 GPU & Hexagon 690 DSP & Min. & Avg. & Max. \\
    \hline
    $230\mu s$ & $20\mu s$ & $0.1\mu s$ & $117\mu s$ & $1209\mu s$ \\
    \hline
   \end{tabular}
   }
  \label{tab:gpu_dsp_overhead}
 \vspace{-2mm}
\end{table}

\begin{figure}[t]
    \centering
    \includegraphics[width=0.48\textwidth]{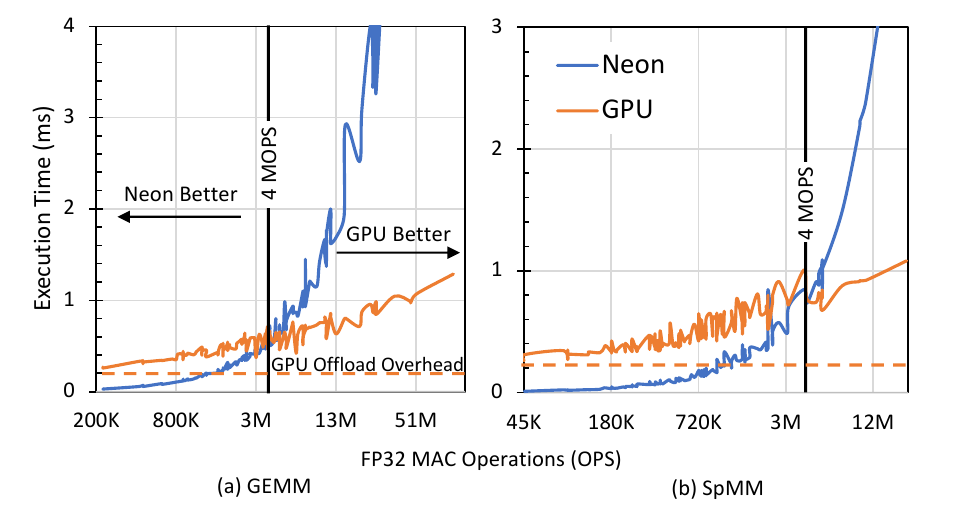}
    \vspace{-6mm}
    \caption{Neon and GPU performance comparison of \libname{XP}'s GEMM and SpMM kernels w.r.t. different operation counts.}
    \vspace{-4mm}
    \label{fig:appprocessor}
\end{figure}

Furthermore, we compare \textit{Neon} and GPU performance of \libname{XP}'s GEMM and SpMM (80\% sparse) kernels for 156 different convolutional layers.
We use 2 OpenCL matrix multiplication libraries~\cite{clblast,clsparse} for GPU and eliminate its memory copy overhead due to the unified memory of mobile SoCs~\cite{unifiedmem}.
Figure~\ref{fig:appprocessor} shows that vector processing outperforms GPU in matrix multiplication with less than 4M FP32 operations despite $96\times$ less throughput.
The primary reason for better \textit{Neon} performance is the lack of offload overhead, thanks to the tight integration with the code.
This overhead is illustrated in the figure by horizontal dash lines.

\section{Conclusion and Future Work}

In this work, we presented \swan{}, the first mobile vector processing benchmark suite with 59 data-parallel kernels from four commonly-used mobile applications.
Using our diverse set of workloads, we analyzed the performance, power, and energy consumption improvement of vectorized kernels, performance bottlenecks of vector processing, limitations of compiler vectorization, and common intensive computation patterns.
In addition, we analyzed the performance scalability of vector processing with wider instructions and more vector execution pipelines.
We discussed the inefficiency of domain-specific acceleration for the fine-grain data-parallel kernels and compared vector processing performance with GPU for different problem sizes.

\swan{} is maintained online on \href{https://github.com/arkhadem/Swan}{GitHub}.
We plan to extend \swan{} in these directions:

\textbf{Other Vector ISA Extensions:}
While Arm Neon is the most widely-used vector architecture of Mobile devices, RISC-V Vector Extension (RVV) \cite{riscv} and Arm Scalable Vector Extension (SVE) \cite{armsve} provide wider registers and sophisticated operations such as Random Memory Accesses.
This is appealing for workloads with irregular memory accesses.
Prior work~\cite{parvec} provides an RVV benchmark suite for Desktop and Server applications.
We aim to equip \swan{} with RVV and SVE implementations of Mobile applications.

\textbf{Android Applications and Java Vector API:}
Java is the primary programming language for Android Application development, which supports vector operations using Java Vector API~\cite{JavaVectorAPI}.
JVBench~\cite{jvbench} provides a benchmark of Java applications and analyzes the performance limitations of Java Just-In-Time Compiler Auto-Vectorization.
However, JVBench borrows data-parallel applications from prior CMP and GPU benchmark suites that are not suitable for vector processing.
We plan to add vectorized Java Applications to \swan{} from various Mobile Application domains such as Social Media, Calendar, E-mail Client, and Navigation.

\textbf{Vectorized Mobile Web Applications}:
While \swan{} contains 12 libraries of the Chromium Project, prior work~\cite{browserenergy1, browserenergy2} show that V8 JavaScript and WebAssembly Engine take a significant portion of the browser's time and energy.
Web Assembly~\cite{WebAssemblyCoreSpecification} is a complementary language to JavaScript for high-performance Web Applications.
V8 JavaScript and WebAssembly Engine supports vector operations through WebAssembly SIMD Proposal~\cite{WebAssemblySIMD}.
We plan to extend the \swan{} benchmark suite with WebAssembly SIMD applications.

\section*{Acknowledgments}

We thank the anonymous reviewers for their suggestions which helped improve this paper.
This work was supported in part by the NSF under the CAREER-1652294 and NSF-1908601 awards, JSPS KAKENHI Grant Number JP22K21284, and the Applications Driving Architectures (ADA) Research Center, a JUMP Center co-sponsored by SRC and DARPA.

\bibliographystyle{IEEEtranS}
\bibliography{main}

\begin{thebibliography}{10}
\providecommand{\url}[1]{#1}
\csname url@samestyle\endcsname
\providecommand{\newblock}{\relax}
\providecommand{\bibinfo}[2]{#2}
\providecommand{\BIBentrySTDinterwordspacing}{\spaceskip=0pt\relax}
\providecommand{\BIBentryALTinterwordstretchfactor}{4}
\providecommand{\BIBentryALTinterwordspacing}{\spaceskip=\fontdimen2\font plus
\BIBentryALTinterwordstretchfactor\fontdimen3\font minus
  \fontdimen4\font\relax}
\providecommand{\BIBforeignlanguage}[2]{{%
\expandafter\ifx\csname l@#1\endcsname\relax
\typeout{** WARNING: IEEEtranS.bst: No hyphenation pattern has been}%
\typeout{** loaded for the language `#1'. Using the pattern for}%
\typeout{** the default language instead.}%
\else
\language=\csname l@#1\endcsname
\fi
#2}}
\providecommand{\BIBdecl}{\relax}
\BIBdecl

\bibitem{android}
\BIBentryALTinterwordspacing
Android - secure and reliable mobile operating system. [Online]. Available:
  \url{https://www.android.com/}
\BIBentrySTDinterwordspacing

\bibitem{pmuevents}
\BIBentryALTinterwordspacing
Arm architecture reference manual for a-profile architecture. [Online].
  Available: \url{https://developer.arm.com/documentation/ddi0487/ja/}
\BIBentrySTDinterwordspacing

\bibitem{cortexa710swopt}
\BIBentryALTinterwordspacing
Arm cortex-a710 core software optimization guide. [Online]. Available:
  \url{https://developer.arm.com/documentation/PJDOC-466751330-14951/latest/}
\BIBentrySTDinterwordspacing

\bibitem{cortexa76swopt}
\BIBentryALTinterwordspacing
Arm cortex-a76 software optimization guide. [Online]. Available:
  \url{https://developer.arm.com/documentation/pjdoc466751330-7215/latest/}
\BIBentrySTDinterwordspacing

\bibitem{chromium}
\BIBentryALTinterwordspacing
Chromium. [Online]. Available: \url{https://www.chromium.org/Home/}
\BIBentrySTDinterwordspacing

\bibitem{multibench}
\BIBentryALTinterwordspacing
``Multibench algorithms and workload datasheets,'' \emph{The Embedded
  Microprocessor Benchmark Consortium}. [Online]. Available:
  \url{https://www.eembc.org/multibench/docs/MultiBench_Algorithms_and_Workload_Datasheets.pdf}
\BIBentrySTDinterwordspacing

\bibitem{pdfium}
\BIBentryALTinterwordspacing
Pdfium. [Online]. Available:
  \url{https://pdfium.googlesource.com/pdfium/+/master/README.md}
\BIBentrySTDinterwordspacing

\bibitem{polybench}
\BIBentryALTinterwordspacing
Polybench. [Online]. Available:
  \url{https://web.cse.ohio-state.edu/~pouchet.2/software/polybench/}
\BIBentrySTDinterwordspacing

\bibitem{riscv}
\BIBentryALTinterwordspacing
riscv-v-spec. [Online]. Available: \url{https://github.com/riscv/riscv-v-spec}
\BIBentrySTDinterwordspacing

\bibitem{webrtc}
\BIBentryALTinterwordspacing
Webrtc. [Online]. Available: \url{https://webrtc.org/}
\BIBentrySTDinterwordspacing

\bibitem{vfp11}
\BIBentryALTinterwordspacing
Arm. Vfp11 vector floating-point coprocessor technical reference manual.
  [Online]. Available:
  \url{https://developer.arm.com/documentation/ddi0274/latest/}
\BIBentrySTDinterwordspacing

\bibitem{jvbench}
\BIBentryALTinterwordspacing
M.~Basso, A.~Ros\`{a}, L.~Omini, and W.~Binder, ``Java vector api: Benchmarking
  and performance analysis,'' in \emph{Proceedings of the 32nd ACM SIGPLAN
  International Conference on Compiler Construction}, ser. CC 2023.\hskip 1em
  plus 0.5em minus 0.4em\relax New York, NY, USA: Association for Computing
  Machinery, 2023, p. 1–12. [Online]. Available:
  \url{https://doi.org/10.1145/3578360.3580265}
\BIBentrySTDinterwordspacing

\bibitem{browserenergy2}
\BIBentryALTinterwordspacing
D.~H. Bui, Y.~Liu, H.~Kim, I.~Shin, and F.~Zhao, ``Rethinking
  energy-performance trade-off in mobile web page loading,'' in
  \emph{Proceedings of the 21st Annual International Conference on Mobile
  Computing and Networking}, ser. MobiCom '15.\hskip 1em plus 0.5em minus
  0.4em\relax New York, NY, USA: Association for Computing Machinery, 2015, p.
  14–26. [Online]. Available: \url{https://doi.org/10.1145/2789168.2790103}
\BIBentrySTDinterwordspacing

\bibitem{parvec}
\BIBentryALTinterwordspacing
J.~M. Cebrian, M.~Jahre, and L.~Natvig, ``Parvec: Vectorizing the parsec
  benchmark suite,'' \emph{Computing}, vol.~97, no.~11, p. 1077–1100, nov
  2015. [Online]. Available: \url{https://doi.org/10.1007/s00607-015-0444-y}
\BIBentrySTDinterwordspacing

\bibitem{rodinia}
S.~Che, M.~Boyer, J.~Meng, D.~Tarjan, J.~W. Sheaffer, S.-H. Lee, and
  K.~Skadron, ``Rodinia: A benchmark suite for heterogeneous computing,'' in
  \emph{2009 IEEE International Symposium on Workload Characterization
  (IISWC)}, 2009, pp. 44--54.

\bibitem{arbench}
S.~Chetoui, R.~Shahi, S.~Abdelaziz, A.~Golas, F.~Hijaz, and S.~Reda, ``Arbench:
  Augmented reality benchmark for mobile devices,'' in \emph{2022 IEEE
  International Symposium on Performance Analysis of Systems and Software
  (ISPASS)}, 2022, pp. 242--244.

\bibitem{mevbench}
J.~Clemons, H.~Zhu, S.~Savarese, and T.~Austin, ``Mevbench: A mobile computer
  vision benchmarking suite,'' in \emph{2011 IEEE International Symposium on
  Workload Characterization (IISWC)}, 2011, pp. 91--102.

\bibitem{clsparse}
\BIBentryALTinterwordspacing
clMathLibraries. A software library containing sparse functions written in
  opencl. [Online]. Available:
  \url{https://github.com/clMathLibraries/clSPARSE/}
\BIBentrySTDinterwordspacing

\bibitem{mmxadoption}
\BIBentryALTinterwordspacing
T.~N. Y.~T. Company. (1994) Makers unveil pcs with intel's mmx chip. [Online].
  Available:
  \url{https://archive.nytimes.com/www.nytimes.com/library/cyber/week/010997intel.html}
\BIBentrySTDinterwordspacing

\bibitem{simpleperf}
\BIBentryALTinterwordspacing
A.~Developers. Simpleperf. [Online]. Available:
  \url{https://developer.android.com/ndk/guides/simpleperf}
\BIBentrySTDinterwordspacing

\bibitem{vectorarchitecture}
R.~Espasa, M.~Valero, and J.~E. Smith, ``Vector architectures: past, present
  and future,'' in \emph{Proceedings of the 12th international conference on
  Supercomputing}, 1998, pp. 425--432.

\bibitem{coremark}
S.~Gal-On and M.~Levy, ``Exploring coremark a benchmark maximizing simplicity
  and efficacy,'' \emph{The Embedded Microprocessor Benchmark Consortium},
  2012.

\bibitem{ultrasparc}
D.~Greenley, J.~Bauman, D.~Chang, D.~Chen, R.~Eltejaein, P.~Ferolito, P.~Fu,
  R.~Garner, D.~Greenhill, H.~Grewal, K.~Holdbrook, B.~Kim, L.~Kohn, H.~Kwan,
  M.~Levitt, G.~Maturana, D.~Mrazek, C.~Narasimhaiah, K.~Normoyle, N.~Parveen,
  P.~Patel, A.~Prabhu, M.~Tremblay, M.~Wong, L.~Yang, K.~Yarlagadda, R.~Yu,
  R.~Yung, and G.~Zyner, ``Ultrasparc: the next generation superscalar 64-bit
  sparc,'' in \emph{Digest of Papers. COMPCON'95. Technologies for the
  Information Superhighway}, 1995, pp. 442--451.

\bibitem{dynamorio}
B.~Hawkins, B.~Demsky, D.~Bruening, and Q.~Zhao, ``Optimizing binary
  translation of dynamically generated code,'' in \emph{Proceedings of the 13th
  Annual IEEE/ACM International Symposium on Code Generation and Optimization},
  ser. CGO '15.\hskip 1em plus 0.5em minus 0.4em\relax USA: IEEE Computer
  Society, 2015, p. 68–78.

\bibitem{topvisited}
\BIBentryALTinterwordspacing
HTMLSTRIP. Alexa top 1000 most visited websites. [Online]. Available:
  \url{https://www.htmlstrip.com/alexa-top-1000-most-visited-websites}
\BIBentrySTDinterwordspacing

\bibitem{WebAssemblySIMD}
\BIBentryALTinterwordspacing
------. Simd proposal for webassembly. [Online]. Available:
  \url{https://github.com/WebAssembly/simd}
\BIBentrySTDinterwordspacing

\bibitem{janapa2022mlperf}
V.~Janapa~Reddi, D.~Kanter, P.~Mattson, J.~Duke, T.~Nguyen, R.~Chukka,
  K.~Shiring, K.-S. Tan, M.~Charlebois, W.~Chou \emph{et~al.}, ``Mlperf mobile
  inference benchmark: An industry-standard open-source machine learning
  benchmark for on-device ai,'' \emph{Proceedings of Machine Learning and
  Systems}, vol.~4, pp. 352--369, 2022.

\bibitem{ramulator}
Y.~Kim, W.~Yang, and O.~Mutlu, ``Ramulator: A fast and extensible dram
  simulator,'' \emph{IEEE Computer Architecture Letters}, vol.~15, no.~1, pp.
  45--49, 2016.

\bibitem{parisc}
R.~Lee, ``Realtime mpeg video via software decompression on a pa-risc
  processor,'' in \emph{Digest of Papers. COMPCON'95. Technologies for the
  Information Superhighway}, 1995, pp. 186--192.

\bibitem{geekbench}
\BIBentryALTinterwordspacing
W.~Lee, J.~Lee, B.~K. Park, and R.~Y.~C. Kim, ``Microarchitectural
  characterization on a mobile workload,'' \emph{Applied Sciences}, vol.~11,
  no.~3, 2021. [Online]. Available:
  \url{https://www.mdpi.com/2076-3417/11/3/1225}
\BIBentrySTDinterwordspacing

\bibitem{llvmcodebook}
B.~C. Lopes and R.~Auler, \emph{Getting started with LLVM core
  libraries}.\hskip 1em plus 0.5em minus 0.4em\relax Packt Publishing Ltd,
  2014.

\bibitem{vectortransformations}
S.~Maleki, Y.~Gao, M.~J. Garzar´n, T.~Wong, and D.~A. Padua, ``An evaluation
  of vectorizing compilers,'' in \emph{2011 International Conference on
  Parallel Architectures and Compilation Techniques}, 2011, pp. 372--382.

\bibitem{vcomputebench}
N.~Mammeri and B.~Juurlink, ``Vcomputebench: A vulkan benchmark suite for gpgpu
  on mobile and embedded gpus,'' in \emph{2018 IEEE International Symposium on
  Workload Characterization (IISWC)}, 2018, pp. 25--35.

\bibitem{vectorbench}
\BIBentryALTinterwordspacing
C.~Mendis, A.~Jain, P.~Jain, and S.~Amarasinghe, ``Revec: Program rejuvenation
  through revectorization,'' in \emph{Proceedings of the 28th International
  Conference on Compiler Construction}, ser. CC 2019.\hskip 1em plus 0.5em
  minus 0.4em\relax New York, NY, USA: Association for Computing Machinery,
  2019, p. 29–41. [Online]. Available:
  \url{https://doi.org/10.1145/3302516.3307357}
\BIBentrySTDinterwordspacing

\bibitem{clblast}
\BIBentryALTinterwordspacing
C.~Nugteren, ``Clblast: A tuned opencl blas library,'' in \emph{Proceedings of
  the International Workshop on OpenCL}, ser. IWOCL '18.\hskip 1em plus 0.5em
  minus 0.4em\relax New York, NY, USA: Association for Computing Machinery,
  2018. [Online]. Available: \url{https://doi.org/10.1145/3204919.3204924}
\BIBentrySTDinterwordspacing

\bibitem{JavaVectorAPI}
\BIBentryALTinterwordspacing
OpenJDK. Jep 426: Vector api (fourth incubator). [Online]. Available:
  \url{https://openjdk.org/jeps/426}
\BIBentrySTDinterwordspacing

\bibitem{browserenergy1}
N.~Peters, S.~Park, S.~Chakraborty, B.~Meurer, H.~Payer, and D.~Clifford, ``Web
  browser workload characterization for power management on hmp platforms,'' in
  \emph{2016 International Conference on Hardware/Software Codesign and System
  Synthesis (CODES+ISSS)}, 2016, pp. 1--10.

\bibitem{cscstar}
\BIBentryALTinterwordspacing
C.~J. Purcell, ``The control data star-100: Performance measurements,'' in
  \emph{Proceedings of the May 6-10, 1974, National Computer Conference and
  Exposition}, ser. AFIPS '74.\hskip 1em plus 0.5em minus 0.4em\relax New York,
  NY, USA: Association for Computing Machinery, 1974, p. 385–387. [Online].
  Available: \url{https://doi.org/10.1145/1500175.1500257}
\BIBentrySTDinterwordspacing

\bibitem{unifiedmem}
\BIBentryALTinterwordspacing
Qualcomm. Heterogeneous memory management. [Online]. Available:
  \url{https://developer.qualcomm.com/software/heterogeneous-compute-sdk/app-notes/buffers}
\BIBentrySTDinterwordspacing

\bibitem{rivec}
\BIBentryALTinterwordspacing
C.~Ram\'{\i}rez, C.~A. Hern\'{a}ndez, O.~Palomar, O.~Unsal, M.~A. Ram\'{\i}rez,
  and A.~Cristal, ``A risc-v simulator and benchmark suite for designing and
  evaluating vector architectures,'' \emph{ACM Trans. Archit. Code Optim.},
  vol.~17, no.~4, nov 2020. [Online]. Available:
  \url{https://doi.org/10.1145/3422667}
\BIBentrySTDinterwordspacing

\bibitem{WebAssemblyCoreSpecification}
\BIBentryALTinterwordspacing
A.~Rossberg, ``Webassembly core specification.'' [Online]. Available:
  \url{https://www.w3.org/TR/wasm-core-1/}
\BIBentrySTDinterwordspacing

\bibitem{cray1}
R.~M. Russell, ``The cray-1 computer system,'' \emph{Communications of the
  ACM}, vol.~21, no.~1, pp. 63--72, 1978.

\bibitem{armsve}
N.~Stephens, S.~Biles, M.~Boettcher, J.~Eapen, M.~Eyole, G.~Gabrielli,
  M.~Horsnell, G.~Magklis, A.~Martinez, N.~Premillieu, A.~Reid, A.~Rico, and
  P.~Walker, ``The arm scalable vector extension,'' \emph{IEEE Micro}, vol.~37,
  no.~2, pp. 26--39, 2017.

\bibitem{parboil}
J.~A. Stratton, C.~Rodrigues, I.-J. Sung, N.~Obeid, L.-W. Chang, N.~Anssari,
  G.~D. Liu, and W.-m.~W. Hwu, ``Parboil: A revised benchmark suite for
  scientific and commercial throughput computing,'' \emph{Center for Reliable
  and High-Performance Computing}, vol. 127, p.~27, 2012.

\bibitem{tiasc}
W.~Watson, ``The ti asc: a highly modular and flexible super computer
  architecture,'' in \emph{Proceedings of the December 5-7, 1972, fall joint
  computer conference, part I}, 1972, pp. 221--228.

\bibitem{topdown}
A.~Yasin, ``A top-down method for performance analysis and counters
  architecture,'' in \emph{2014 IEEE International Symposium on Performance
  Analysis of Systems and Software (ISPASS)}, 2014, pp. 35--44.

\bibitem{wibench}
Q.~Zheng, Y.~Chen, R.~Dreslinski, C.~Chakrabarti, A.~Anastasopoulos, S.~Mahlke,
  and T.~Mudge, ``Wibench: An open source kernel suite for benchmarking
  wireless systems,'' in \emph{2013 IEEE International Symposium on Workload
  Characterization (IISWC)}, 2013, pp. 123--132.

\end{thebibliography}

\end{document}